\documentclass[lettersize,journal]{IEEEtran}
\usepackage{amsmath,amssymb,amsfonts}
\usepackage{amsthm}
\usepackage{graphicx}
\usepackage{algorithm}
\usepackage{algorithmicx}
\usepackage{array}
\usepackage{textcomp}
\usepackage[noend]{algpseudocode}
\usepackage{textcomp}
\usepackage{xcolor}
\usepackage{xspace}
\usepackage{pifont}
\usepackage{subfigure}
\usepackage{balance}
\newtheorem{definition}{Definition}
\newtheorem{example}{Example}
\usepackage{enumitem}
\usepackage{ulem}
\usepackage{soul}
\usepackage{stfloats}
\usepackage{url}
\usepackage{verbatim}
\usepackage{cite}
\newtheorem{theorem}{Theorem}
\newtheorem{lemma}{Lemma}
\def\BibTeX{{\rm B\kern-.05em{\sc i\kern-.025em b}\kern-.08em
		T\kern-.1667em\lower.7ex\hbox{E}\kern-.125emX}}
\newcommand{\mokplex}{maximal antagonistic $k$-plex\xspace}
\newcommand{\mokplexes}{maximal antagonistic $k$-plexes\xspace}
\newcommand{\stitle}[1]{\noindent{\bf #1}}

\setlist[itemize]{nosep,label=\textbf{-},leftmargin=*}
\setlist[enumerate]{nosep,label=\textit{(\alph*)},leftmargin=*}

\begin{document}

\title{Efficient Antagonistic $k$-plex Enumeration in Signed Graphs}

%\author{IEEE Publication Technology,~\IEEEmembership{Staff,~IEEE,}
\author{Lantian Xu, Rong-Hua Li, Dong Wen, Qiangqiang Dai, Guoren Wang, Lu Qin
        % <-this % stops a space
\thanks{Lantian Xu and Lu Qin are with the Australian
 	Artificial Intelligence Institute, University of Technology Sydney, Sydney, Australia. }% <-this % stops a space
\thanks{Rong-Hua Li, Qiangqiang Dai and Guoren Wang are with Beijing
	Institute of Technology (BIT), Beijing, China. }% <-this % stops a space
\thanks{Dong Wen is with the University of
	New South Wales, Sydney, Australia. }% <-this % stops a space
\thanks{Manuscript received April 19, 2021; revised August 16, 2021.}}

% The paper headers
\markboth{Journal of \LaTeX\ Class Files,~Vol.~14, No.~8, August~2021}%
{Shell \MakeLowercase{\textit{et al.}}: A Sample Article Using IEEEtran.cls for IEEE Journals}

%\IEEEpubid{0000--0000/00\$00.00~\copyright~2021 IEEE}
% Remember, if you use this you must call \IEEEpubidadjcol in the second
% column for its text to clear the IEEEpubid mark.

\maketitle

\begin{abstract}
A signed graph is a graph where each edge receives a sign, positive or negative. The signed graph model has been used in many real applications, such as protein complex discovery and social network analysis. Finding cohesive subgraphs in signed graphs is a fundamental problem. A $k$-plex is a common model for cohesive subgraphs in which every vertex is adjacent to all
but at most $k$ vertices within the subgraph.
%The mining of cohesive subgraphs on signed graphs can be used in many fields such as protein complex detection, phrases discovery, etc. $k$-plex is an important model in cohesive subgraphs search. A $k$-plex is a subgraph in which every vertex is adjacent to all but at most $k$ vertices within the subgraph. Existing algorithms based on $k$-plex models mainly focus on unsigned networks, and these methods cannot be directly applied to signed graphs due to the positive and negative attributes of edges. 
In this paper, we propose the model of size-constrained antagonistic $k$-plex in a signed graph. The proposed model guarantees that the resulting subgraph is a $k$-plex and can be divided into two sub-$k$-plexes, both of which have positive inner edges and negative outer edges. This paper aims to identify all maximal antagonistic $k$-plexes in a signed graph. Through rigorous analysis, we show that the problem is NP-Hardness. We propose a novel framework for maximal antagonistic $k$-plexes utilizing set enumeration. Efficiency is improved through pivot pruning and early termination based on the color bound. Preprocessing techniques based on degree and dichromatic graphs effectively narrow the search space before enumeration. Extensive experiments on real-world datasets demonstrate our algorithm's efficiency, effectiveness, and scalability. 
\end{abstract}


\section{Introduction}

\IEEEPARstart{S}{igned} graphs serve as effective tools for representing the polarity of relationships between entities, employing positive and negative symbols to denote the associations between the respective vertices. These graphs find applications in diverse domains, such as capturing friend-foe relationships in social networks\cite{DBLP:journals/tasm/Chen13}, expressing support-dissent opinions within opinion networks\cite{DBLP:conf/www/KunegisLB09}, characterizing trust-distrust relationships in trust networks\cite{DBLP:conf/sdm/GiatsidisCMTV14}, and depicting activation-inhibition dynamics in protein-protein interaction networks\cite{DBLP:journals/tcbb/Ou-YangDZ15}. 

\begin{figure}[t!]\vspace*{-0.3cm}
	\centering
	\includegraphics[height=3cm]{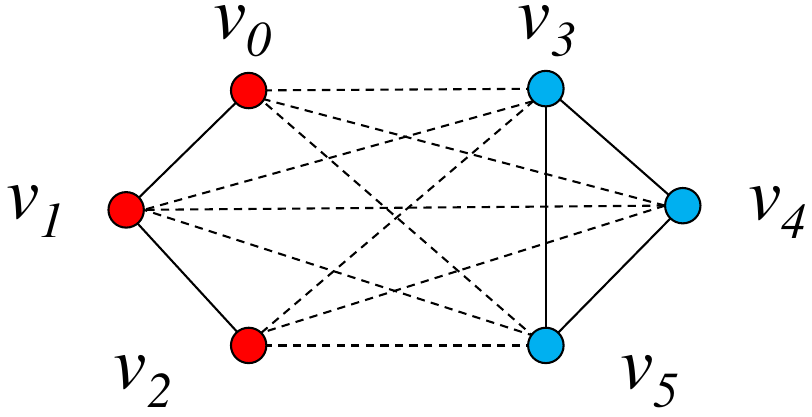}
	\vspace*{-0.3cm} \caption{Balanced graph}
	\label{fig:Balanced Graph} 
\end{figure}

Structural balance theory is an essential and foundational theory in the analysis of signed graphs. According to this theory, a signed graph denoted as $G$ is considered balanced if it can be partitioned into two distinct subgraphs, where edges within each subgraph are positive, and the edges connecting vertices from different subgraphs are negative\cite{harary1953notion}. That is, “The friend (resp. enemy) of my friend(resp. enemy) is my friend, the friend (resp. enemy) of my enemy (resp. friend) is my enemy”. We can find the antagonistic sub-communities as the localized effect of social balance theory\cite{DBLP:journals/datamine/GaoLLP16}. Consider the graph $G$ shown in Figure \ref{fig:Balanced Graph}, solid (resp. dashed) lines represent positive (resp. negative) edges. $G$ can be divide into two parts, one part is $\left\{v_{0},v_{1},v_{2}\right\}$, the other part is $\left\{v_{3},v_{4},v_{5}\right\}$. Although there are no egdes between $v_{0}$ and $v_{2}$, $v_{0}$ and $v_{2}$ have the common friend $v_{1}$ and common enemies $v_{3},v_{4}$ and $v_{5}$. According to the structural balance theory, $v_{0}$ and $v_{2}$ can also be regarded as friends.

Some cohesive subgraph models in signed graphs have also been investigated in the literature. \cite{chen2020efficient} proposed the definition of balanced clique and gave the maximal balanced clique search algorithm. Based on this, \cite{yaocomputing} further proposed a search algorithm for a maximum balanced clique. Due to data noise, clique, where vertices are pairwise connected, can rarely appear in real data\cite{DBLP:conf/kdd/ConteFMPT17,DBLP:journals/ior/BalasundaramBH11}. It may be too strict to use clique to find cohesive subgraphs. In order to more accurately mine cohesive subgraphs in signed graphs, a relaxed model of cliques is expected.

%A $k$-plex is a vertex set that is nearly a clique but each vertex of the $k$-plex is allowed to have $k$ missing adjacent vertices in this vertex set, $k$ being a positive integer. A $k$-plex is said to be maximal if there does not exist any other $k$-plex that contains it. To effectively use $k$-plexes in detecting communities, a fundamental problem is to enumerate all maximal $k$-plexes\cite{DBLP:conf/aaai/ZhouXGXJ20}, which often arises in a large number of real-world applications, such as community detection in social networks\cite{balasundaram2011clique}, identifying protein complexes in PPI networks\cite{luo2009core}, and serving as an alternative to cliques in biochemistry\cite{doyle2005robust}.

The $k$-plex is a crucial cohesive subgraph model adopted extensively in graph analysis. However, its relevant definitions and algorithms for signed networks are less studied. We introduce the maximal antagonistic $k$-plex model for cohesive subgraphs in signed graphs to address this gap and draw inspiration from structural balance theory \cite{harary1953notion}. Formally, given a signed network $G$, a \mokplex $C$ is a maximal subgraph of $G$ such that (1) $C$ is $k$-plex without considering the symbols of edges. (2) $C$ is antagonistic, i.e., $C$ can be divided into two parts such that the edges in the same part are positive, and the edges connecting two parts are negative. To derive large $k$-plexes, we specify the minimum size of each antagonistic part as a parameter $t$.
% the plex by setting a parameter $t$. The two antagonistic parts each contain at least $t$ vertices. 
This definition incorporates both density and balance considerations. We prove that the \mokplex problem is NP-hard. The primary objective of our paper is to develop efficient algorithms for enumerating all \mokplex subgraphs within a given signed network.

\stitle{Applications.} We show several applications of maximal antagonistic $k$-plex enumeration as follows.

\noindent\underline{\textit{- Antagonistic group detection.}} People often share or argue with each other on social networks such as Facebook and Quora. Users with agreements can be represented by positive edges, while negative edges can represent those with disagreements. Consequently, social network data can be modeled as a signed graph \cite{kumar2018community, xiao2020searching, bonchi2019discovering}. The main objective of the antagonistic $k$-plex is to identify two opposing opinion groups in a social network, where users within each group tend to agree with one another and disagree with users from the other group. By computing \mokplex, we can efficiently discover user groups with contrasting perspectives. These groups are typically tightly interconnected and hold significant influence within the network.

% \noindent\underline{\textit{- Protein complex detection.}} In the analysis of protein-protein interaction (PPI) networks, the signed graph model enables a deeper understanding of activation-inhibition relationships among proteins \cite{suratanee2014characterizing, yim2018annotating}. Protein complexes, characterized by closely interacting proteins, can be viewed as proteomes with strong positive interactions within the complex and robust negative interactions between different complexes. Therefore, identifying antagonistic $k$-plex in signed PPI networks can serve as a valuable method for detecting protein complexes.

\noindent\underline{\textit{- Protein complex detection.}} To reconstruct the signaling pathway from the PPI network, we need the signatures of the PPIs, which represent whether the interaction has a positive or negative effect. A signed graph can represent activation-inhibition relationships among proteins in the PPI network\cite{suratanee2014characterizing, yim2018annotating}. Protein complexes can be defined as a group of proteins in which there is a dense population of positively interacting (i.e., activating) proteins, and a dense population of negatively interacting (i.e., inhibiting)\cite{ou2015detecting,yaocomputing}. Therefore, identifying antagonistic $k$-plex in signed PPI networks can serve as a valuable method for detecting protein complexes.

\noindent\underline{\textit{- Synonyms and antonymic phrases discovery.}} Signed graphs provide a natural way to represent synonym and antonym relationships between words \cite{miller1995wordnet}. Utilizing maximal antagonistic $k$-plex, we can discover synonym groups that are antonymous with each other, such as \{sonant, voiced, 
loud, hard\} and \{surd, soft,voiceless, unvoiced\}. These discovered clusters can be further utilized in applications such as automatic question
generation \cite{kumar2019cross} and semantic expansion \cite{krishnan2018leveraging}.

\stitle{Contributions.} We make the following contributions:

\noindent\ul{\textit{- A new $k$-plex model for signed graphs.}} We formalize the antagonistic plex model in signed networks based on the structural balance theory. As far as we know, the paper is the first work considering the maximal antagonistic $k$-plex in signed networks. We  prove the NP-Hardness of the problem.

\noindent\ul{\textit{- A new framework tailored for maximal antagonistic plex enumeration.}} In conjunction with the definition of antagonistic $k$-plex, we introduce antagonistic $k$-plex expansion conditions. We employ set enumeration techniques to correctly discover of all eligible maximal antagonistic $k$-plexes, facilitating a comprehensive exploration of the solution space.

\noindent\ul{\textit{- Novel optimization strategies to improve the enumeration performance.}} We present early termination conditions to speed up set enumeration. Using the pivot technique, we reduce the search branch during enumeration. Moreover, the color bound in the signed graph allows premature termination of unpromising searches, improving the efficiency of identifying maximal antagonistic $k$-plex in signed networks.

\noindent\ul{\textit{- Preprocessing before set enumeration.}} In the preprocessing phase, we employ degree pruning in the signed graph, followed by converting it into a dichromatic graph and applying the plex pruning rules. We demonstrate that plexes infeasible in the dichromatic graph cannot form antagonistic $k$-plex in the signed graph. Utilizing this insight, we propose a method to remove unqualified vertices from the dichromatic graph, thereby reducing the size of the signed graph before set enumeration.

\noindent\ul{\textit{- Extensive performance studies on real datasets.}} We perform comprehensive experiments to assess the performance of our proposed algorithms on various real datasets. The results demonstrate that our optimized approach achieves significantly faster execution, nearly two orders of magnitude quicker than the basic algorithm.

\stitle{Outline.} Section 2 provides preliminaries including the definition of antagonistic $k$-plex model and problem statement. Section 3 introduces the basic algorithm. Section 4 shows several optimization techniques. Section 5 reports the results of experimental studies. Section 6 shows some related works. Section 7 concludes our paper.

\vspace*{-0.1cm}
\section{Preliminaries}
\vspace*{-0.1cm}

In this paper, we focus on the undirected signed graph $G = (V, E)$, where $V$ is the set of vertices and $E$ is the set of signed edges. Each edge in $E$ is either positive or negative. We use $E^{+}$ and $E^{-}$ to denote the positive and negative edges, respectively, where $E = E^+ \cup E^-$. There are no multi-edges in $G$.
% So we also can denote the signed graph as $G = (V, E^{+}, E^{-})$. 
We denote the number of vertices and the number of edges by $n$ and $m$, respectively, i.e., $n=|V|$ and $m=|E|=|E^{+}|+|E^{-}|$. Let $N_{G}^{+}(v)$ represent the positive neighbors of $v$, i.e., $N_{G}^{+}(v)=\left\{u|(v,u)\in E^{+}\right\}$. Let $N_{G}^{-}(v)$ represent the negative neighbors of $v$, i.e., $N_{G}^{-}(v)=\left\{u|(v,u)\in E^{-}\right\}$. We use $d_{G}^{+}=|N_{G}^{+}(v)|$ and $d_{G}^{-}=|N_{G}^{-}(v)|$ to denote the positive degree and negative degree of $v$, respectively. We denote $N_{G}(v)=N_{G}^{+}(v) \cup N_{G}^{-}(v)$ by $N_{G}(v)$ and $d_{G}(v)=d_{G}^{+}(v) + d_{G}^{-}(v)$ by $d_{G}(v)$.
Let $N_{G}^{2}(v)$ be the 2-hop neighbor of $v$, i.e., $N_{G}^{2}(v) = \{u | N_{G}(u) \cap N_G(v) \neq \emptyset, u \not\in N_G(v)\}$. 

We denote $v$'s positive neighbors' positive neighbors as $N_{G}^{++}(v)$. Let $w_{1},w_{2}$...$w_{n}$ be the vertices in $N_{G}^{+}(v)$, $N_{G}^{++}(v)=\bigcup_{1}^{n} N_{G}^{+}(w_{i})$. In the same way, we define $N_{G}^{+-}(v)$, $N_{G}^{-+}(v)$, and $N_{G}^{--}(v)$, which represent $v$'s positive neighbors' negative neighbors, $v$'s negative neighbors' positive neighbors, and $v$'s negative neighbors' negative neighbors respectively.
Given two vertices $u,v$, if $(u,v)\in E^{+}$, we think they are friends and belong to the same group. If $(u,v)\in E^{-}$, we think they are opponents and belong to the antagonistic groups. 
Further, the 2-hop neighbors of $v$ also can be put into two groups, i.e., the candidate group of $v$'s friends and the candidate group of $v$'s opponents. We use $N_{G}^{2+}(v )$ to denote $v$'s 2-hop neighbors which may join the same group with $v$, i.e., $N_{G}^{2+}(v )= N_{G}^{++}(v )\cup N_{G}^{--}(v )$. We use $N_{G}^{2-}(v )$ to denote $v$'s 2-hop neighbors which may join the different group with $v$, i.e., $N_{G}^{2-}(v )= N_{G}^{-+}(v )\cup N_{G}^{+-}(v )$. Note that a vertex can be contained in both $N_{G}^{2+}(v )$ and $N_{G}^{2-}(v )$.
%  For example, $\exists x\in N_{G}^{+}(v)$ and $\exists y\in N_{G}^{+}(v)$, $(u,x)\in E^{-}$ and $(u,y)\in E^{+}$, in this case $u\in N_{G}^{2+}(v )$ and $u\in N_{G}^{2-}(v )$. 

% \vspace*{-0.2cm}

% \begin{definition}[Antagonistic Graph\cite{chen2020efficient}]
	% 	A signed graph $G = (V, E^{+}, E^{-})$ is antagonistic if $V$ can be split into two subsets $C_{L}$ and $C_{R}$, s.t.$u\in C_{L},v\in C_{R} \rightarrow (u,v)\notin E$ or $(u,v)\in E^{-}$, $u,v\in C_{L}$ or $u,v\in C_{R} \rightarrow (u,v)\notin E$ or $(u,v)\in E^{+}$.
	% \end{definition}\label{def:antagonistic Community}
% \textcolor{blue}{balanced vs antagonistic?\cite{harary1953notion}}

\begin{definition}[Antagonistic Graph\cite{chen2020efficient}]
	A signed graph $G = (V, E^{+}, E^{-})$ is antagonistic if $V$ can be split into two subsets $C_{L}$ and $C_{R}$, s.t., $\forall (u,v) \in E^+ \to u,v \in C_L$ or $u,v\in C_{R}$; and
	$ \forall (u,v)\in E^{-} \to u\in C_{L},v\in C_{R}$ or $u \in C_R, v \in C_L$.
\end{definition}\label{def:antagonisticCommunity}

A $k$-plex is a subgraph in which every vertex connects to at least $s-k$ vertices in the subgraph where $s$ is the number of vertices in the subgraph \cite{seidman1978graph}. It is clear that any subgraph of a $k$-plex is also a $k$-plex. 

% \vspace*{-0.1cm}

\begin{definition}[Maximal Antagonistic $k$-plex]\label{def:mokplex}
	Given a signed network $G = (V, E^{+}, E^{-})$, a maximal antagonistic $k$-plex $C$ is a maximal subgraph of $G$ that satisfies: (1) $C$ is $k$-plex; and (2) $C$ is antagonistic.
\end{definition}

% \begin{definition}[Maximal Antagonistic $k$-plex]\label{def:mokplex}
	% 	Given a signed network $G = (V, E^{+}, E^{-})$, a maximal antagonistic $k$-plex $C$ is a maximal subgraph of $G$ that satisfies:
	% 	\begin{enumerate}
		% 		\item 
		% 		$C$ is $k$-plex.i.e. $ \forall u \in C, d_{C}(u)\ge |C|-k$;
		
		% 		\item
		% 		$C$ is antagonistic, i.e, it can be split into two plex $C_{L}$ and $C_{R}$ which are also $k$-plex, s.t. $\forall u,v \in C_{L}$ or $\forall u,v \in C_{R} \to \left(u,v\right) \in E^{+}$ or $\left(u,v\right)$ not exist, and $\forall u \in C_{L},\forall v \in C_{R}$ or $\forall u \in C_{R},\forall v \in C_{L} \to \left(u,v\right) \in E^{-}$ or $\left(u,v\right)$ not exist.
		% 	\end{enumerate}
	
	% \end{definition}

\vspace*{-0.1cm}

To guarantee the cohesiveness of resulting $k$-plexes, several existing works \cite{DBLP:conf/aaai/ZhouXGXJ20,conte2018d2k} only consider $k$-plexes with at least $2k-1$ vertices. We also follow this setting.

% However, if we do not limit the diameter of the plex, the vertices in the plex may be not close enough. Some vertices may not be adjacent and not share a common neighbor. So we require a maximum diameter of 2. It is equivalent to saying that every two vertices in the community are directly linked or at least share a common neighbor. Lemma \ref{lemma:d} shows the relationship between $t$ and $k$ in a plex with a diameter of at least 2.

\begin{lemma}[Bounded Diameter \cite{seidman1978graph}]\label{lemma:d}
	Given a $k$-plex $C$ with the size of $s$, $C$ is connected and the diameter of $C$ is at most $2$ if $s \geq 2k-1$.
\end{lemma}

The bounded diameter guarantees that any two vertices in the $k$-plex are close and at least share a common neighbor. 
We allow users to control the size of resulting $k$-plexes by setting a size threshold of at least $2k-1$ for $C_L$ and $C_R$. 
The research problem is formally presented as follows.

% In this paper, we aim to enumerate all \mokplex in a given signed network. The diameter of the plex is at least 2. Users can control the size and the slackness of the return result. We add a size constraint to $C_{L}$ and $C_{R}$, so the user can control the size of the returned results. Users can also use $k$ to control the slackness of the antagonistic plex. We formalize the studied problem as follows: 

\stitle{Problem statement.} Given a signed network $G$ and two integers $k$ and $t\ge 2k-1$, we aim to compute all maximal connected antagonistic $k$-plex $C$ in $G$ s.t. $|C_{L}|\ge t$ and $|C_{R}|\ge t$.

%\textcolor{red}{(is the result connected?)}

For each resulting \mokplex, we guarantee the cohesiveness of $C_L$, $C_R$, and the whole subgraph. Note that the number of all vertices in each resulting subgraph is at least $4k-2$, and the diameter is also bounded by $2$.

\begin{figure}[t!]\vspace*{-0.3cm}
	\centering
	\includegraphics[height=5cm]{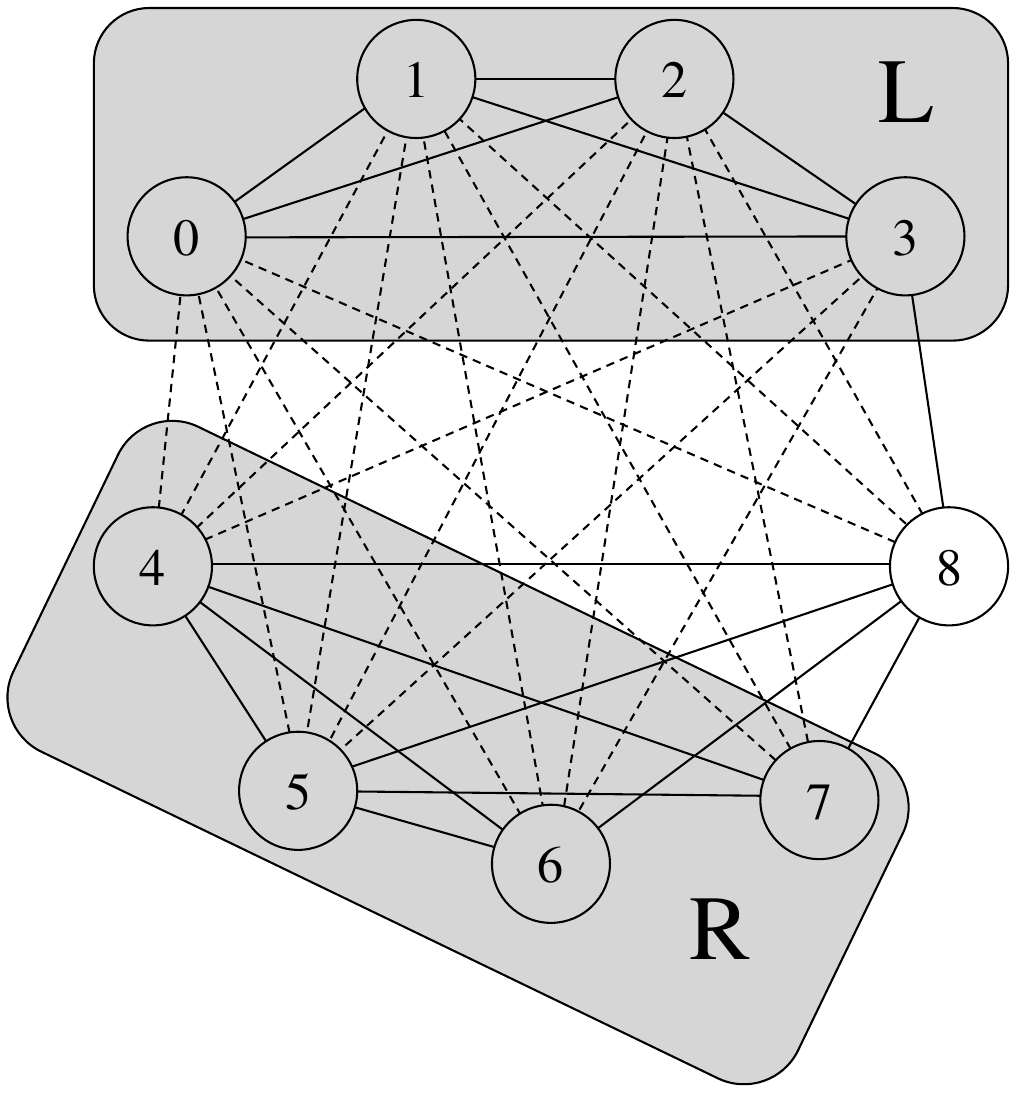}
	\vspace*{-0.3cm} \caption{Maximal antagonistic $k$-plex}
	\label{fig:exampleplex} 
\end{figure}

\begin{example}
	Figure \ref{fig:exampleplex} is a signed network. The solid/dashed lines denoted positive/negative edges. If we set $t=4$ and $k=2$, there is a \mokplex in this graph. This \mokplex $C$ can be divide into two parts, $C_{L}=\left\{v_{0},v_{1},v_{2},v_{3}\right\}$ and $C_{R}=\left\{v_{4},v_{5},v_{6},v_{7}\right\}$, where
	vertices in $C_{L}$ and $C_{R}$ are marked with different markings.  $C_{L}$ is complete as any two vertices in $C_{L}$ have an positive edge. It is clear that $C_{L}$ is a positive $k$-plex. In  $C_{R}$, only $v_{6}$ and $v_{7}$ are not adjacent. So $C_{R}$ is still a positive $k$-plex. Consider the negative edges between $C_{L}$ and $C_{R}$, only $v_{3}$ and $v_{7}$ do not have a negative edge. It meets the requirement of $k$-plex. Though $v_{8}$ is adjacent to all the other vertices, it has positive edges with $V_{3}$ and all the vertices in $C_{R}$. That means $v_{8}$ cannot be added into the \mokplex. $C \cup \left\{v_{8}\right\}$ is not qualified for an antagonistic community. Therefore, the plex $C$ is maximal.
\end{example}

\begin{theorem}[Problem Hardness]
	The \mokplex enumeration problem is NP-Hard.
\end{theorem}

\begin{proof}
	It can be proved following the NP-Hardness of maximal $k$-plex enumeration problem\cite{balasundaram2011clique}. According to definition \ref{def:mokplex}, all the \mokplex is $k$-plex, if we treat the positive and negative edges as the same. In other words, \mokplex is a kind of more complex $k$-plex. We can even list all of the $k$-plex first and test each of them to judge if they can be \mokplex. Although in the research problem of this paper we are given a size constraint, finding all eligible \mokplexes cannot be solved in non-deterministic polynomial-time. It is clear that \mokplex is NP-hard.
\end{proof}

\section{A Basic Algorithm}

We first propose a basic algorithm based on existing techniques for maximal $k$-plex enumeration in unsigned networks\cite{wang2022listing,wang2017parallelizing,wu2007parallel} and maximal balanced clique enumeration\cite{chen2020efficient} in signed networks. 
Our framework maintains an antagonistic $k$-plex $C$ by two vertex sets $C=\{C_L,C_R\}$ based on Definition \ref{def:mokplex}. Let $P_{L}$ be the set of candidate vertices that can be added into $C_{L}$, and $P_{R}$ be the set of candidate vertices that can be added into $C_{R}$. In each step, we expand $C$ by adding vertices from $P_{L}$ and $P_{R}$ into $C_{L}$ and $C_{R}$, respectively. When a new vertex is added into $C$, we should update $P_{L}$ and $P_{R}$. Until no more vertices can join $C$, we can stop and output $C_{L}$ and $C_{R}$ as a \mokplex. Moreover, we use $Q_{L}$ and $Q_{R}$ to record candidate vertices that have been processed to avoid outputting duplicate \mokplex. If $Q_L$ or $Q_R$ is not empty, the corresponding antagonistic $k$-plex $C=C_L \cup C_R$ is contained in an earlier result.

% In this step, we also need to maintain all the vertices in $C$ whose degrees are $|C|-k$. For all vertices in $P_{L}(Q_{L})$ and $P_{R}(Q_{R})$, they should meet the requirement \ding{1}\ding{3} and \ding{2}\ding{3} in Lemma \ref{lem:antagonistic $k$-plex expansion conditions} respectively. 

Our basic algorithm for \mokplex enumeration is presented in Algorithm \ref{alg:BAPE}. We process vertices $v_{0},v_{1}, . . . ,v_{n}$ (Line 2) in the ascending order of $min(d_{G}^{+},d_{G}^{-})$. We say a vertex $u$ ranks higher than $v$ if $u$ is in front of $v$ in the order.
For each vertex $v_{i}$, we enumerate all \mokplexes containing $v_{i}$ (Line 2--8). $C_{L}$ and $C_{R}$ are initialized by $v_{i}$ and $\emptyset$, respectively (Line 3). We initialize $P_{L}$ and $P_{R}$ with vertices ranking lower than $v_{i}$ in the candidate set. We initialize $Q_{L}$ and $Q_{R}$ with vertices ranks higher than $v_{i}$ in the candidate set. After initializing these six sets, we invoke procedure \texttt{BAPEUTIL} to enumerate all \mokplexes containing $v_{i}$ (Line 8).

\stitle{Expanding the subgraph.}
% Next, we discuss how to expand an antagonistic $k$-plex. 
Given a $k$-plex $C$ and a vertex $v$,  $S \cup \{v\}$ is also a $k$-plex if and only if (1) $v$ is adjacent to all the vertices in $S$ which satisfies $d_{S}(v)=|S|-k$; and (2) $|N_{G}(v)\cap S| \geq |S|+1-k$. We extend the conditions to the context of signed networks.

\begin{lemma}\label{lem:antagonistic $k$-plex expansion conditions} 
	Given an antagonistic $k$-plex $C=\left\{C_{L},C_{R}\right\}$ and a vertex $v$, $C'=C\cup \left\{v\right\}$ is also an antagonistic $k$-plex if $v$ satisfies \ding{172}\ding{174} or \ding{173}\ding{174} as follows.
	
	\begin{itemize}
		\item [\ding{172}] 
		$\forall  u\in \left\{u\in C_{L}|d_{C}(u)=|C|-k\right\} \rightarrow (u,v)\in E^{+}$;\\$\forall u\in C_{L} \rightarrow (u,v)\notin E^{-}$;\\$\forall  u\in \left\{u\in C_{R}|d_{C}(u)=|C|-k\right\} \rightarrow (u,v)\in E^{-}$;\\$\forall u\in C_{R} \rightarrow (u,v)\notin E^{+}$.
		
		\item [\ding{173}]
		$\forall  u\in \left\{u\in C_{R}|d_{C}(u)=|C|-k\right\} \rightarrow (u,v)\in E^{+}$;\\$\forall u\in C_{R} \rightarrow (u,v)\notin E^{-}$;\\$\forall  u\in \left\{u\in C_{L}|d_{C}(u)=|C|-k\right\} \rightarrow (u,v)\in E^{-}$;\\$\forall u\in C_{L} \rightarrow (u,v)\notin E^{+}$.
		
		\item [\ding{174}]
		$|N_{G}(v) \cap C|\geq |C|+1-k$.
	\end{itemize}
	
	\noindent	
	Specifically, if $v$ satisfies \ding{172}\ding{174}, $C'=\left\{C_{L}\cup \left\{v\right\},C_{R}\right\}$. If $v$ satisfies \ding{173}\ding{174}, $C'=\left\{C_{L},C_{R}\cup \left\{v\right\}\right\}$.
\end{lemma}\label{lemma:antagonistic $k$-plex expansion conditions}

\vspace*{-0.3cm}

\begin{proof}
	According to Definition \ref{def:mokplex}, the whole subgraph is a $k$-plex. So, the new vertex $v$ must satisfy \ding{174}. $v$ must also simultaneously connect all vertices $u$ in $C$ with $d_{v}(u)=|C|-k$. At the same time, $v$ must be attributed to either $C_{L}$ or $C_{R}$. Then, the edges between $v$ and the vertex $u$ with $d_{v}(u)=|C|-k$ must also conform to the antagonistic principle of positive edges between vertices in the same group and negative edges between vertices in different groups. It is also possible that $v$ is connected to other vertices in $C$ with $d_{v}(u)\neq|C|-k$, and then these edges must likewise not violate the antagonistic principle. Therefore, at least one of \ding{172} and \ding{173} is satisfied.
\end{proof}

\begin{algorithm}[t!]
	\footnotesize
	\caption{\texttt{BAPE}$(G=(V,E^{+},E^{-}),k,t)$}
	\label{alg:BAPE} 
	\hspace*{\algorithmicindent} \textbf{Input: }a signed graph $G$,$k$,$t$ \\
	\hspace*{\algorithmicindent} \textbf{Output: }All maximal antagonistic $k$-plex
	\begin{algorithmic}[1]
		\State $Flag \gets  true$ 
		\For{$v_{i}\in \left \{ v_{0},v_{1},...,v_{n-1}\right \}$}
		\State $C_{L}\gets  \left \{v_{i} \right \},C_{R}\gets  \emptyset$
		\State $P_{L}\gets (N_{G}^{+}(v_{i} )\cup N_{G}^{2+}(v_{i} )) \cap \left \{ v_{i+1},...,v_{n-1}\right \}$
		\State $P_{R}\gets (N_{G}^{-}(v_{i} )\cup N_{G}^{2-}(v_{i} )) \cap \left \{ v_{i+1},...,v_{n-1}\right \}$
		\State $Q_{L}\gets (N_{G}^{+}(v_{i} )\cup N_{G}^{2+}(v_{i} )) \cap \left \{ v_{0},...,v_{i-1}\right \}$
		\State $Q_{R}\gets (N_{G}^{-}(v_{i} )\cup N_{G}^{2-}(v_{i} )) \cap \left \{ v_{0},...,v_{i-1}\right \}$
		\State \texttt{BAPEUTIL}$\left ( C_{L},C_{R},P_{L},P_{R},Q_{L},Q_{R},k,t\right )$
		\EndFor
		
		\Procedure{\texttt{BAPEUTIL}}{$C_{L},C_{R},P_{L},P_{R},Q_{L},Q_{R},k,t$}
		\State $P_{L} \gets $\texttt{update}$(P_{L},C_{L},C_{R},k)$
		\State $Q_{L} \gets $\texttt{update}$(Q_{L},C_{L},C_{R},k)$
		\State $P_{R} \gets $\texttt{update}$(P_{R},C_{L},C_{R},k)$
		\State $Q_{R} \gets $\texttt{update}$(Q_{R},C_{L},C_{R},k)$
		%		\State $P_{L}\gets \left \{v\in P_{L}|G [C_{L}\cup C_{R}\cup \left \{v \right \}] \text{ is a $k$-plex} \right \}  $
		%		\State $P_{R}\gets \left \{v\in P_{R}|G [C_{L}\cup C_{R}\cup \left \{v \right \}] \text{ is a $k$-plex} \right \}  $
		%		\State $Q_{L}\gets \left \{v\in Q_{L}|G [C_{L}\cup C_{R}\cup \left \{v \right \}] \text{ is a $k$-plex} \right \}  $
		%		\State $Q_{R}\gets \left \{v\in Q_{R}|G [C_{L}\cup C_{R}\cup \left \{v \right \}] \text{ is a $k$-plex} \right \}  $
		\If{$P_{L}=\emptyset$ and $P_{R}=\emptyset$ and $Q_{L}=\emptyset$ and $Q_{R}=\emptyset$}
		\If{$|C_{L}|\geq t$ and $|C_{R}|\geq t$}
		\State return $C=\left \{C_{L},C_{R} \right\}$
		\EndIf
		\EndIf
		\State $Flag \gets  !Flag$ 
		\If{$Flag$}
		\For{$v \in P_{L}$}
		\State $P_{L}\gets P_{L} \setminus \left\{ v\right\}$
		\State \texttt{BAPEUTIL}$(C_{L}\cup \left\{ v\right\},C_{R}, P_{L}, P_{R}, Q_{L}, Q_{R})$
		\State $Q_{L}\gets Q_{L} \cup \left\{ v\right\}$
		\EndFor
		\For{$v \in P_{R}$}
		\State $P_{R}\gets P_{R} \setminus \left\{ v\right\}$
		\State \texttt{BAPEUTIL}$(C_{L},C_{R}\cup \left\{ v\right\}, P_{L}, P_{R}, Q_{L}, Q_{R})$
		\State $Q_{R}\gets Q_{R} \cup \left\{ v\right\}$
		\EndFor
		\Else
		\State Line 23--26; Line 19--22
		\EndIf
		\EndProcedure
		
	\end{algorithmic}
	
\end{algorithm}

Based on Lemma \ref{lem:antagonistic $k$-plex expansion conditions}, we can locate a set of candidate vertices that can be added to the current antagonistic $k$-plex. When $C_L$ or $C_R$ expands, we need to refine the candidate set.
Algorithm \ref{alg:update} presents the \texttt{update} procedure given the new $C_L$ and $C_R$. The input set $X$ can be $P_{L}$, $P_{R}$, $Q_{L}$ or $Q_{R}$. 
Supported by the \texttt{update} (Algorithm \ref{alg:update}) procedure, \texttt{BAPEUtil} performs the \mokplex enumeration based on the given six sets. If $P_{L}$, $P_{R}$, $Q_{L}$ and $Q_{R}$ are empty (Line 14), the current antagonistic plex $C=\left\{C_{L},C_{R}\right\}$ cannot be enlarged and is a \mokplex. \texttt{BAPEUtil} further checks whether $C_{L}$ and $C_{R}$ satisfy the size constraint. If so, it outputs the current antagonistic plex (Line 15--16). Otherwise, \texttt{BAPEUtil} adds a vertex $v$ from $P_{L}$ to $C_{L}$ and recursively invokes itself for further expansion (Line 21). When $v$ is processed, $v$ is removed from $P_{L}$ and added in $Q_{L}$ (Line 22). Similar processing steps are applied on vertices in $P_{R}$ (Line 23--26). In order to balance the size of $C_L$ and $C_R$ when searching, we first expand $C_R$ in the next call by judging the $Flag$ (Line 28).

\begin{algorithm}[t!]
	\footnotesize
	\caption{\texttt{update}$(X,C_{L},C_{R},k)$}
	\label{alg:update} 
	\begin{algorithmic}[1]
		\State $Xnew=\emptyset$
		\If{$X$ is $P_{L}$ or $Q_{L}$}
		\For{$v$ in $X$}
		\If{$v$ meet requirement \ding{172} and \ding{174}}
		\State $Xnew \gets Xnew \cup \left\{v\right\}$
		\EndIf
		\EndFor
		\EndIf
		\If{$X$ is $P_{R}$ or $Q_{R}$}
		\For{$v$ in $X$}
		\If{$v$ meet requirement \ding{173} and \ding{174}}
		\State $Xnew \gets Xnew \cup \left\{v\right\}$
		\EndIf
		\EndFor
		\EndIf
		\State \textbf{Return} $Xnew$
	\end{algorithmic}
	
\end{algorithm}

\begin{figure}[t!]\vspace*{-0cm}
	\centering
	\includegraphics[height=4.5cm]{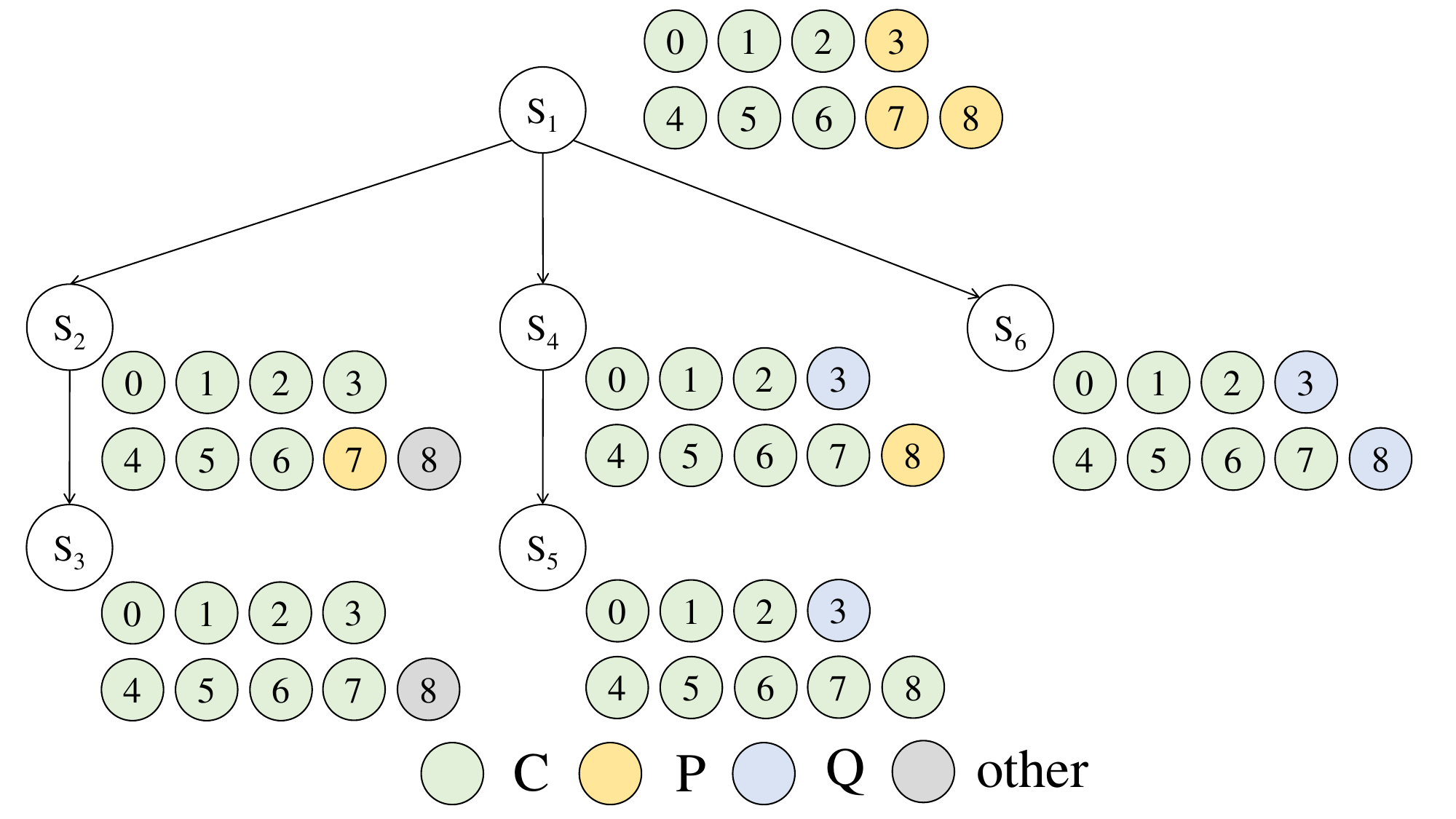}
	\vspace*{-0.2cm} \caption{Running example}
	\label{fig:exampleplexenum} \vspace*{0.2cm}
\end{figure}

\begin{theorem}[Correctness]
	Algorithm \ref{alg:BAPE} finds all \mokplexes correctly without redundancy.
\end{theorem}
\vspace*{-0.5cm}
\begin{proof}
	We show the correctness of Algorithm \ref{alg:BAPE} from three aspects: (1) The antagonistic $k$-plex output in Algorithm \ref{alg:BAPE} is maximal. Assume that an antagonistic $k$-plex $C$ output in Line
	16 is not maximal. The candidate set updated by Algorithm \ref{alg:update} would not be empty simultaneously. For a vertex $v$ can be added into $C$ and make $C\cup \left\{v\right\}$ be a larger plex, if $v$ is behind the current enumeration vertex $u$, $v$ should be stored in $P_{L}$ or $P_{R}$. $C$ cannot be output in this round. Otherwise, $v$ should be stored in $Q_{L}$ or $Q_{R}$. It cannot satisfy the conditions in Line 14, and $C$ would not be output, either. (2) Algorithm \ref{alg:BAPE} will output all the qualified \mokplex. In Line 2, Algorithm \ref{alg:BAPE} visits each vertex $v_{i}$. Based on the recursive structure of \texttt{BAPE}, all the \mokplex containing $v_{i}$ are explored. (3) Algorithm \ref{alg:BAPE} will not output the same  \mokplex more than once. If the current enumeration vertex is $v_{i}$ and the current plex has been output in the previous round, the current plex must contain a vertex $v_{j}(j<i)$. In the previous round, $v_{j}$ would be added into $Q_{L}$ or $Q_{L}$. Therefore, in the current round, $Q_{L}$ or $Q_{L}$ will not be empty, and the plex will not be output. Combining all the above three aspects, the correctness of Algorithm \ref{alg:BAPE} is proved.
\end{proof}

\begin{example}
	The enumeration procedure of \texttt{BAPE} can be illustrated as a search tree. Figure \ref{fig:exampleplexenum} shows part of the search tree when we conduct the \texttt{BAPE} ($k=2,t=4$) on $G$ in Figure \ref{fig:exampleplex} through \texttt{BAPE}. $S_{1}$, $S_{2}$, . . . represent different search states during the enumeration. At $S_{1}$, we assume that we have an antagonistic $k$-plex $C=\left\{C_{L}=\left\{v_{0},v_{1},v_{2}\right\},C_{R}=\left\{v_{4},v_{5},v_{6}\right\}\right\}$ at this state. First, we add $v_{3}$ from $P_{L}$ to $C_{L}$ in $S_{2}$. Due to the negative edge between $v_{3}$ and $v_{8}$, we should delete $v_{8}$ from $P_{R}$. Then, $v_{7}$ is added into $C_{R}$. At $S_{3}$, there are no vertices in $P$ or $Q$. So $C=\left\{C_{L}=\left\{v_{0},v_{1},v_{2},v_{3}\right\},C_{R}=\left\{v_{4},v_{5},v_{6},v_{7}\right\}\right\}$ can be output as a \mokplex. In $S_{4}$, we add $v_{7}$ into $C_{R}$ first. Now, $v_{3}$ is in $Q_{L}$. Though $v_{8}$ is added into $C_{R}$ in $S_{5}$, the current plex does not qualify due to the size of $C_{L}$ being too small. Similarly, no plex will be output in $S_{6}$.
\end{example}

%\textbf{Complexity analysis of Algorithm \ref{alg:BAPE}}
We analyze the complexity of our basic approach and start from the time complexity of \texttt{update} as follows.

\begin{theorem}\label{lem:update}
	The Algorithm \texttt{update} $(X,C_{L},C_{R},k)$ runs in $O(|C|^{2}+|X|(|C_{L}|+|C_{R}|))$ time.
\end{theorem}

\begin{proof}
	Algorithm \ref{alg:update} can be divided into two parts. First, according to the expansion conditions, we should find all the vertices $u$ in $C_{L}$ and $C_{R}$ such that $d_{C}(u)=|C|-k$. This
	step can be finished within running time $O(|C|^{2})$. Second, for each vertex $v$ in $X$, we should test its adjacency to all vertices in $C$, and we can compute $|N_{G}(v)\cap C|$ simultaneously. This step can be done in $O(|X|(|C_{L}|+|C_{R}|))$.
\end{proof}

% \vspace*{-0.4cm}

% \textcolor{red}{\sout{\texttt{update} is invoked four times in \texttt{BAPEUTIL}.}}

% In Algorithm \ref{alg:BAPE}, the Algorithm \ref{alg:update} needs to process four sets, i.e., $P_{L}$, $Q_{L}$, $P_{R}$ and $Q_{R}$, so it is called four times in each iteration. 
%The whole running time of update process is $O(|C|^{2}+(|P_{L}|+|Q_{L}|+|P_{R}|+|Q_{R}|)(|C_{L}|+|C_{R}|))$ in each iteration.

We consider the size of $P_{L}$, $Q_{L}$, $P_{R}$ and $Q_{R}$ which are the inputs of \texttt{BAPEUTIL}. The total size of the four sets is bounded by $N_{G}^{2}(v_{i})$. Let $\Delta$ be the maximum degree of $v\in G$. We have $|N_{G}^{2}(v_{i})|\leq \Delta^{2}$. Based on Theorem \ref{lem:update}, \texttt{update} runs in $O(\Delta^{4})$. For every $v_{i}$ in $G$, \texttt{BAPE} invokes \texttt{BAPEUTIL} once.

\begin{theorem}
	The worst-case time complexity of Algorithm \ref{alg:BAPE} is $O(n\Delta^{4}2^{\Delta^{2}})$.
\end{theorem}

\vspace*{-0.4cm}

\begin{proof}
	Consider the case where the input subgraph of the function \texttt{BAPEUTIL} is an antagonistic $k$-plex itself. Due to the hereditary property
	of $k$-plex, the input subgraph of \texttt{BAPEUTIL} is $k$-plex in each iteration. Therefore, the \texttt{update} process cannot remove vertices from $P$. In this case, \texttt{BAPEUTIL} will be called $|P_{jL}|+|P_{jR}|$ times in $j$-th iteration, $P_{jL}$ and $P_{jR}$ are the symbol of $P_{L}$ and $P_{R}$ in $j$-th iteration. Due to the property of set enumeration, the total times is $O(2^{|P_{L}|+|P_{R}|})\leq O(2^{\Delta^{2}})$. Combining with Theorem \ref{lem:update}, the running
	time of the update can be assumed as $O(\Delta^{4})$. Hence, the
	whole running time of Algorithm \ref{alg:BAPE} in the worst case is $O(\Delta^{4}\sum_{i=0}^{n}2^{\Delta^{2}})\leq O(n\Delta^{4}2^{\Delta^{2}})$.
\end{proof}

It is worth mentioning that although we do not introduce pruning on degree in this section, degree-based VR pruning (see Section \ref{po} for details) is included in our experiments with the baseline algorithm(\texttt{BAPE}).

\vspace*{-0.2cm}
\section{OPTIMIZATION}
\vspace*{-0.1cm}

% Although Algorithm \ref{alg:BAPE} can solve the \mokplex enumeration problem, the efficiency of Algorithm \ref{alg:BAPE} is not efficient enough. 
% In this section, we present several novel optimization techniques to improve the efficiency of the basic approach.

% \vspace*{-0.2cm}
\subsection{Enumeration Optimization}

We present several novel optimization techniques to improve the efficiency of the basic approach. We improve the key recursive procedure \texttt{BAPEUTIL}. The updated algorithm is called \texttt{SAPEUTIL}, and the pseudocode is presented in Algorithm \ref{alg:SAPEUTIL}.

\stitle{Pivoting-based pruning technique.} We utilize the pivot technique to prune the unnecessary branches in the search tree of Algorithm \ref{alg:BAPE}.

\begin{lemma}[$k$-plex pivoting\cite{conte2018d2k}]\label{lemma:k-plexpivoting}
	In an undirected unsigned graph, let $C$ be a $k$-plex, $P$ is the candidate set, i.e., $P=\left\{v\notin K| K\cup \left\{v\right\} \text{is a $k$-plex}\right\}$, and $u$ is a vertex in $P$. Any maximal $k$-plex containing $C$ contains either $u$, a non-neighbor of $u$, or a neighbor $v$ of $u$ such that $v$ and $u$ have a common non-neighbor in $K$.
\end{lemma}

As shown in lemma \ref{lemma:k-plexpivoting}, the existing pivot method only works for unsigned graphs. We extend it to the antagonistic $k$-plex in signed graphs.

\begin{lemma}[Antagonistic $k$-plex pivoting]\label{lemma:k-plexpivotsign}
	In a signed graph, let $C=\left\{C_{L},C_{R}\right\}$ be a $k$-plex, candidate set be $P=\left\{v\notin C| C\cup \left\{v\right\} \text{is a $k$-plex}\right\}$, and $u$ be a vertex in $P$. Any maximal $k$-plex containing $C_{L}$ contains either $u$, a non-neighbor of $u$, or a neighbor $v$ of $u$ such that $v$ and $u$ have a common non-neighbor in $C_{L}$. Any maximal $k$-plex containing $C_{R}$ contains either $u$, a non-neighbor of $u$, or a neighbor $v$ of $u$ such that $v$ and $u$ have a common non-neighbor in $C_{R}$. 
\end{lemma}

\begin{proof}
	In the signed graph, if we want to select a vertex from the candidate set and put it into the current antagonistic plex, the vertex we choose should be able to form two plexes simultaneously. Specifically, if we want to move $v$ from $P_{L}$ to $C_{L}$, we must meet two conditions. First, considering only the positive side, $C_{L}\cup \left\{v\right\}$ is a $k$-plex. Second, considering only $v$'s negative edges and the positive edges in $C_{R}$, $C_{R}\cup \left\{v\right\}$ is also a $k$-plex.
\end{proof}

\begin{algorithm}[t!] 
	\footnotesize
	\caption{\texttt{SAPEUTIL}$(C_{L},C_{R},P_{L},P_{R},Q_{L},Q_{R},k,t)$}
	\label{alg:SAPEUTIL} 
	\begin{algorithmic}[1]
		\State $P_{L} \gets $\texttt{update}$(P_{L},C_{L},C_{R},k)$
		\State $Q_{L} \gets $\texttt{update}$(Q_{L},C_{L},C_{R},k)$
		\State $P_{R} \gets $\texttt{update}$(P_{R},C_{L},C_{R},k)$
		\State $Q_{R} \gets $\texttt{update}$(Q_{R},C_{L},C_{R},k)$
		%		\State $P_{L}\gets \left \{v\in P_{L}|G [C_{L}\cup C_{R}\cup \left \{v \right \}] \text{ is a $k$-plex} \right \}  $
		%		\State $P_{R}\gets \left \{v\in P_{R}|G [C_{L}\cup C_{R}\cup \left \{v \right \}] \text{ is a $k$-plex} \right \}  $
		%		\State $Q_{L}\gets \left \{v\in Q_{L}|G [C_{L}\cup C_{R}\cup \left \{v \right \}] \text{ is a $k$-plex} \right \}  $
		%		\State $Q_{R}\gets \left \{v\in Q_{R}|G [C_{L}\cup C_{R}\cup \left \{v \right \}] \text{ is a $k$-plex} \right \}  $
		\If{$P_{L}=\emptyset$ and $P_{R}=\emptyset$ and $Q_{L}=\emptyset$ and $Q_{R}=\emptyset$}
		\If{$|C_{L}|\geq t$ and $|C_{R}|\geq t$}
		\State return $C=\left \{C_{L},C_{R} \right\}$
		\EndIf
		\EndIf
		\If{$|C_{L}|+|P_{L}| < t$ or $|C_{R}|+|P_{R}| < t$}
		return
		\EndIf
		\State Partition vertices of $P_{L},P_{R},P_{L}\cup P_{R}$ by greedy coloring heuristic. They colornums are $cd^{L}$,$cd^{R}$,$cd^{A}$ respectively.
		\If{$cd^{L}<t$ or $cd^{R}<t$ or $cd^{A}<2t$}
		continue
		\EndIf
		\State $Flag \gets  !Flag$ 
		\If{$Flag$}
		\State choose a pivot $u$ from $P_{L}\cup Q_{L}$
		\State $A_{L}=\left \{ c \in P_{L}\cap N_{G}\left ( u \right )|C_{L} \setminus (N_{G}^{+} ( u  ) \cup N_{G}^{+} (c)  ) = \emptyset \right \}$
		\State $B_{L}=\left \{ c \in P_{L}\cap N_{G}\left ( u \right )|C_{R} \setminus (N_{G}^{-} ( u  ) \cup N_{G}^{-} (c)  )= \emptyset \right \}$
		\For{$v \in P_L \setminus N_G(u) \cap A_L \cap B_L$}
		\State $P_{L}\gets P_{L} \setminus \left\{ v\right\}$
		\If{$cd^{L}_{v}<t$ or $cd^{A}_{v}<2t$}  continue
		\EndIf
		
		\State \texttt{SAPEUTIL}$(C_{L}\cup \left\{ v\right\},C_{R}, P_{L}, P_{R}, Q_{L}, Q_{R})$
		\State $Q_{L}\gets Q_{L} \cup \left\{ v\right\}$
		\EndFor
		\State $A_{R}=\left \{ c \in P_{R}\cap N_{G}\left ( u \right )|C_{R} \setminus (N_{G}^{+} ( u  ) \cup N_{G}^{+} (c)  )= \emptyset \right \}$
		\State $B_{R}=\left \{ c \in P_{R}\cap N_{G}\left ( u \right )|C_{L} \setminus (N_{G}^{-} ( u  ) \cup N_{G}^{-} (c)  )= \emptyset \right \}$
		\For{$v \in P_R \setminus N_G(u) \cap A_R \cap B_R$}
		\State $P_{R}\gets P_{R} \setminus \left\{ v\right\}$
		\If{$cd^{R}_{v}<t$ or $cd^{A}_{v}<2t$}
		continue
		\EndIf
		
		\State \texttt{SAPEUTIL}$(C_{L},C_{R}\cup \left\{ v\right\}, P_{L}, P_{R}, Q_{L}, Q_{R})$
		\State $Q_{R}\gets Q_{R} \cup \left\{ v\right\}$
		\EndFor
		\Else
		\State choose a pivot $u$ from $P_{R} \cup Q_{R}$
		\State Line 21--27; Line 14--20
		\EndIf

	\end{algorithmic}
	
\end{algorithm}

% \stitle{\underline{Algorithm Implementation.}} Now we modify Algorithm \ref{alg:BAPE} according to lemma \ref{lemma:k-plexpivotsign}. 

To apply Lemma \ref{lemma:k-plexpivotsign}, we choose $u$ as a pivot. In order to maximize the effectiveness of this cut, we adopt the philosophy of Chen et al \cite{chen2020efficient,chen2020efficient}, and select a vertex $u$ with maximum $|N^{+(-)}(u)\cap P_{L}|+|N^{-(+)}(u)\cap P_{R}|$.
%[describe how to choose a pivot]} 
%
%\textcolor{blue}{[you can directly describe how you do it in \texttt{SAPEUTIL}, instead of describing how to modify \texttt{BAPEUTIL}].}
Then, we compute all neighbors of $u$ in the candidate set $P_{L}$ with no common non-neighbor with $u$ in $C_{L}$. The result is denoted as $A_{L}$. We also compute all neighbors of $u$ in $P_{L}$ that have no common non-neighbor with $u$ in $C_{R}$. The result is denoted as $B_{L}$. The pivoting technique replaces Line 19 in Algorithm \ref{alg:BAPE} with the following operation:
\begin{equation*}
\text{For } v \in P_L \setminus N_G(u) \cap A_L \cap B_L
\end{equation*}

% For $v \in (P_{L} \setminus N_{G}\left ( u \right )) \cup (N_{G}\left ( u \right )-A_{L}\cap B_{L})$ 

Similarly, we compute all neighbors of $u$ in the candidate set $P_{R}$ with no common non-neighbor with $u$ in $C_{R}$. The result is denoted as $A_{R}$. We also compute all neighbors of $u$ in $P_{R}$ that have no common non-neighbor with $u$ in $C_{L}$. The result is denoted as $B_{R}$. We replace Line 23 in Algorithm \ref{alg:BAPE} with the following operation:

\begin{equation*}
\text{For } v \in P_R \setminus N_G(u) \cap A_R \cap B_R
\end{equation*}

% For  $v \in (P_{R}\setminus N_{G}\left ( u \right )) \cup (N_{G}\left ( u \right )-A_{R}\cap B_{R})$

By pivoting, we reduce the number of recursions and the search scope. In pivoting-based pruning, we must identify if an edge exists between a vertex in $P_L$ (or $P_R$) and a vertex in $C_L$ (or $C_R$). Therefore, the pruned candidate set can be computed in $O((|P_{L}|+|P_{R}|)(|C_{L}|+|C_{R}|))$.

% traverse the candidate set $P_{L}(P_{R})$ and determine the adjacency of each vertex in P with the vertices in $C_{L}$ and $C_{R}$. So this pruning can be finished in $O((|P_{L}|+|P_{R}|)(|C_{L}|+|C_{R}|))$.

\stitle{Early termination.} 
% We consider a method that we can terminate the search early in Algorithm \ref{alg:BAPE}. 
For an antagonistic $k$-plex $C=\left\{C_{L},C_{R}\right\}$, if all vertices in $P_{L}$ and $P_{R}$ can be moved to $C$, the maximal possible size of $C_{L}$ and $C_{R}$ for the final
\mokplex are $|C_{L}| + |P_{L}|$ and $|C_{R}| + |P_{R}|$, respectively. Therefore, we can terminate the search if $|C_{L}| + |P_{L}|<k$ or $|C_{R}| + |P_{R}|<k$. We apply the rule in Line 8 of Algorithm \ref{alg:SAPEUTIL}, which can be done in constant time.

\stitle{Colorbound-based pruning technique.} We extend the color bound method for $k$-plex search in ordinary graphs and propose color-bound-based pruning on signed graphs.

\begin{figure}[t!]\vspace*{0.3cm}
\centering
\includegraphics[height=3cm]{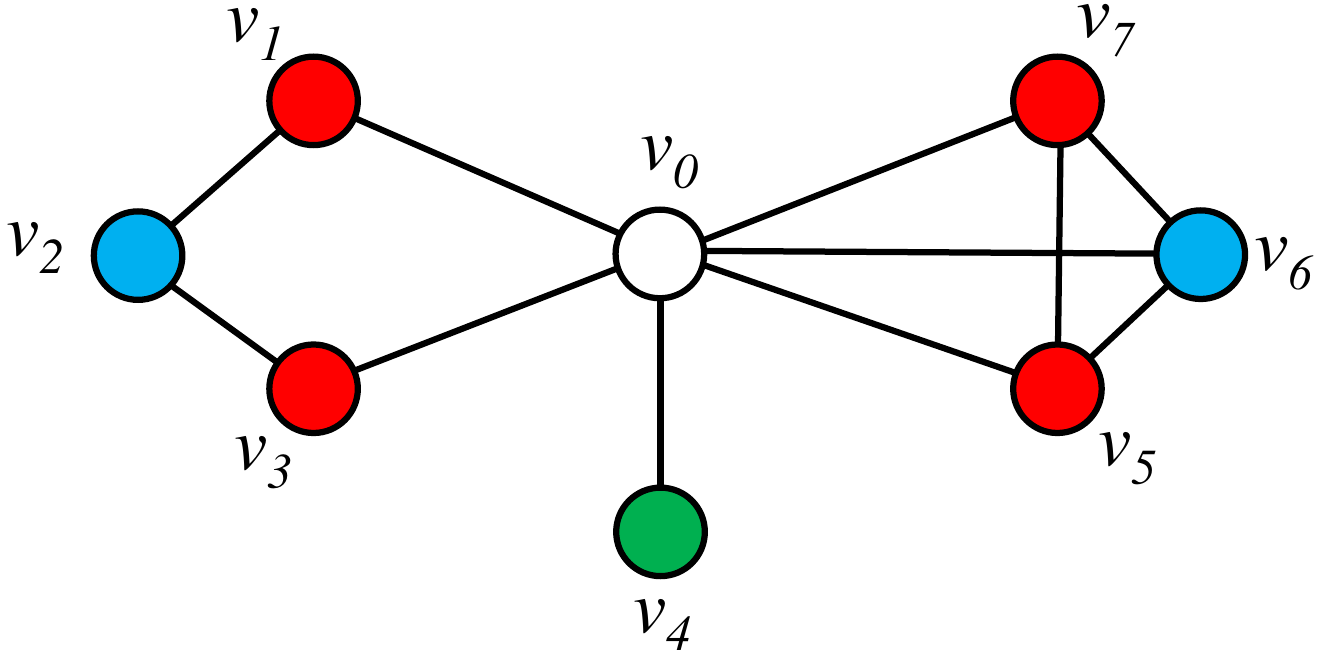}
\vspace*{-0.1cm} \caption{Color bound example}
\label{fig:Color Bound example} 
\end{figure}

\begin{lemma}[Color-bound\cite{zhou2021improving}]\label{lemma:color-bound}
Given a graph $G=(V,E)$, if $V$ can be partitioned into $c$ disjoint independent sets $I_{1}...I_{c}$, then the upper bound of the size of maximum
$k$-plex in $G$ is $\sum_{i=1}^{c}min\left \{ |I_{i}|,k \right \}$, a.k.a. color-bound.
\end{lemma}

According to the size limit in the problem definition, we guarantee three plexes satisfy the size constraint simultaneously in a signed graph. The three plexes are the plex in $C_{L}\cup P_{L}$ only containing positive edges, the plex in $C_{R}\cup P_{R}$ only containing positive edges and the whole plex containing all edges. We apply the lemma \ref{lemma:color-bound} for three sets to prune unnecessary search space.

\begin{definition}[Colornum in Signed Graph]\label{def:colornum in signed graph}
In an intermediate state of the enumeration of \mokplex, $P_{L}$ can be partitioned into $c$ disjoint independent sets $pl_{1},...pl_{c}$ (only containing positive edges), $P_{R}$ can be partitioned into $c$ disjoint independent sets $pr_{1},...pr_{c}$ (only containing positive edges), $P$ can be can be partitioned into $c$ disjoint independent sets $a_{1},...a_{c}$ (containing positive and negative edges). Colornums in signed graph can be computed as follows, when $L=C_{L}\cup P_{L}$, $R=C_{R}\cup P_{R}$ and $A=C_{L}\cup P_{L}\cup C_{R}\cup P_{R}$.
%are the upper bound of the sizes of maximum $k$-plex of $L=C_{L}\cup P_{L}$, $R=C_{R}\cup P_{R}$ and $A=C_{L}\cup P_{L}\cup C_{R}\cup P_{R}$.

$cd^{L} \gets \sum_{j=1}^{c}min\left ( |pl_{j}|,k \right )+|C_{L}|$

$cd^{R} \gets \sum_{j=1}^{c}min\left ( |pr_{j}|,k \right )+|C_{R}|$

$cd^{A} \gets \sum_{j=1}^{c}min\left ( |a_{j}|,k \right )+|C_{L}\cup C_{R}|$

\end{definition}

\begin{lemma}
\label{lemma:colornum in signed graph}
Colornums in signed graph are the upper bound of the sizes of maximum $k$-plex of $L$, $R$ and $A$.
\end{lemma}
%\textcolor{red}{(the lemma reads like a definition. Should be better to first define what is colornum, then give a lemma says the colornum is an upper bound)}

\begin{proof}
According to Definition \ref{def:mokplex}, there are three $k$-plexes in an antagonistic $k$-plex. Two small $k$-plexes that consider only positive edges, and a large $k$-plex that considers both positive and negative edges. Then, in the signed graph, we can compute three colornum upper bounds for the three $k$-plexes. Any eligible antagonistic $k$-plex should satisfy all three upper bounds simultaneously.
\end{proof}

\begin{example}
Figure \ref{fig:Color Bound example} shows the example of color bound reduction. In a intermediate state of the enumeration, $C_{L}=\left\{v_{0}\right\}$, $P_{L}=\left\{v_{1},v_{2},v_{3},v_{4},v_{5},v_{6},v_{7}\right\}$. According to lemma \ref{lemma:colornum in signed graph}, the vertices in $P_{L}$ are divided into three independent sets and colored with different colors, i.e., $pl_{1}=\left\{v_{1},v_{3},v_{4},v_{5}\right\}$, $pl_{2}=\left\{v_{2},v_{6}\right\}$ and $pl_{3}=\left\{v_{7}\right\}$. Suppose $k=2$, then the colornum of $P_{L}$ is 5.
\end{example}

\stitle{\underline{Algorithm Implementation.}} We use the rule of color bound in Algorithm \ref{alg:SAPEUTIL} (Lines 9--10). 
After each update of the candidate set, we greedily color the vertices of all remaining candidate sets.
Then, we calculate colornums of $P_{L}$, $P_{R}$ and $A$ based on Lemma \ref{lemma:colornum in signed graph}. We can skip the iteration if the upper bound of the maximum plex that these candidate sets produce is less than our requirement.

\begin{lemma}[Color-degree of One Vertex\cite{zhou2021improving}]\label{lemma:color-bound of one vertex}
In an intermediate state of the enumeration with a growing $k$-plex $C$, a candidate set $P$, assume $I_{1}...I_{c}$ is a coloring
of $P$. For $u\in P$, the size of $k$-plex $S$ that $u\in S$ and $C\subseteq S$ is bounded by $\sum_{j=1,u\notin I_{j}}^{c}min\left ( |I_{j}|\cap N_{G}(u),k \right )+\left ( k-|(C\setminus N_{G}(u)| \right )+|C|$.
\end{lemma}

Similarly, if we want to add a vertex $u$ into $C_{L}$ or $C_{R}$, we can test the color-degree of $u$ first.

\begin{theorem}[Color-degree in Signed Graph]\label{lemma:color-degree in signed graph}
In an intermediate state of the enumeration with two growing $k$-plex $C_{L}$ and $C_{R}$, two candidate sets $P_{L}$ and $P_{R}$, assume $pl_{j}$ is a coloring of $P_{L}$, assume $pr_{j}$ is a coloring of $P_{R}$, assume $a_{j}$ is a coloring of $A_{L}$. For $v\in P_{L}(P_{R})$, the size of $k$-plex $S$ that $v\in S$ and $C\subseteq S$ is bounded by following:

%	$cd^{L}_{v} \gets \sum_{j=1,u\notin pl_{j}}^{c}min\left ( |pl_{j}|\cap N_{G}^{+}(v),k \right )+\left ( k-|C_{L}\setminus N_{v}^{+}(v)| \right )+|C_{L}|$

$cd^{L}_{v} \gets F\left(pl_{j},N_{G}^{+}(v)\right)+\left ( k-|C_{L}\setminus N_{v}^{+}(v)| \right )+|C_{L}|$

%	$cd^{R}_{v} \gets \sum_{j=1,u\notin pr_{j}}^{c}min\left ( |pr_{j}|\cap N_{G}^{+}(v),k \right )+\left ( k-|C_{R}\setminus N_{v}^{+}(v)| \right )+|C_{R}|$

$cd^{R}_{v} \gets F\left(pr_{j},N_{G}^{+}(v)\right)+\left ( k-|C_{R}\setminus N_{v}^{+}(v)| \right )+|C_{R}|$

%	$cd^{A}_{v} \gets \sum_{j=1,u\notin a_{j}}^{c}min\left ( |a_{j}|\cap N_{G}(v),k \right )+\left ( k-|(C_{L}\cup C_{R})\setminus N_{G}(v)| \right )+|C_{L}\cup C_{R}|$

$cd^{A}_{v} \gets F\left(a_{j},N_{G}(v)\right)+\left ( k-|(C_{L}\cup C_{R})\setminus N_{G}(v)| \right )+|C|$

In all the above expression, we set $F\left(X,Y\right)=\sum_{j=1,u\notin X}^{c}min\left ( |X|\cap Y,k \right )$.
\end{theorem}

\begin{proof}
Similarly to Lemma \ref{lemma:colornum in signed graph}, we can similarly compute three corresponding color-degrees on three $k$-plexes for each candidate vertex $v$. Any eligible $v$ should satisfy the constraints on all three of its color-degrees simultaneously.
\end{proof}

\begin{algorithm}[t!] 
\footnotesize
\caption{\texttt{VertexReduction}$(G=(V,E^{+},E^{-}),k,t)$}
\label{alg:VertexReduction}

\begin{algorithmic}[1]
	\While{$\exists v \in V,d_{G}^{+}\left(v\right)<t-k$ or $d_{G}^{-}\left(v\right)<t-k+1$ or $d_{G}\left(v\right)<2t-k$}
	\For{$u\in N_{G}^{+}\left(v\right)$}
	\State $d_{G}^{+}\left(u\right)-=1$
	\EndFor
	\For{$u\in N_{G}^{-}\left(v\right)$}
	\State $d_{G}^{-}\left(u\right)-=1$
	\EndFor
	\State $G\gets G \setminus v$
	\EndWhile
	\State \textbf{Return} $G$

\end{algorithmic}

\end{algorithm}

\begin{example}
In Figure \ref{fig:Color Bound example}, suppose $k=2$ and try to add $v_{7}$ into $C_{L}$. We can see $min\left ( |pl_{1}|\cap N_{G}^{+}(v_{7}),k \right )=1$, $min\left ( |pl_{2}|\cap N_{G}^{+}(v_{7}),k \right )=1$, $\left ( k-|(C_{L}\cup C_{R})\setminus N_{G}(v)| \right )=0$, so $cd^{L}_{v_{7}}=5$. If $t>5$, $v_{7}$ can be skipped.
\end{example}

We use this rule in Algorithm \ref{alg:SAPEUTIL} (Lines 18 and 25). When we finish updateing, we test if the candidate set can build a qualified plex by Lemma \ref{lemma:colornum in signed graph}. This step can be done in $O(\Delta|P|)$. Then, every time we try to add a vertex $v$ int $C_{L}(C_{R})$, we can calculate color-degree of $v$ based on lemma \ref{lemma:color-degree in signed graph}. If the upper bound of the maximum plex that $v$ can build in this iteration is less than our requirement, we can skip $v$. For all vertices in $P$, we need to iterate over its neighbor vertices. Therefore, the time is also $O(\Delta|P|)$.

\subsection{Preprocessing Optimization}
\label{po}

In preprocessing optimization, we aim to remove worthless vertices and edges not contained in any \mokplex according to the constraints.

\stitle{Vertex Reduction (VR).} We first consider the neighbors of each vertex. According to the definition \ref{def:mokplex}, for each vertex $v$ in a \mokplex $C$, the number of $v$'s neighbors is at least $|C|-k$. Given that $|C|$ is at least $2t$, we have $d_{G}(v)\geq 2t-k$.

$C_{L}$ and $C_{R}$ also should be $k$-plex. The edges between vertices in $C_{L}$ ($C_{R}$) are positive, which implies the positive degree is at least $t-k$, i.e., $d_{G}^{+}(v)\geq t-k$. Further, we consider the negative degree. The negative edge only exists between vertices from $C_{L}$ and $C_{R}$, respectively. Each vertex $v$ in $C_{L}$ can build a plex with $C_{R}$ if we only consider the negative edges from $v$ and the positive edge among $C_{R}$. Similarly, each vertex $v$ in $C_{R}$ can build a plex with $C_{L}$ if we only consider the negative edges from $v$ and the positive edge among $C_{L}$. That means the negative degree should be at least $t-k+1$, i.e., $d_{G}^{-}(v)\geq t-k+1$.

Then, we use the three constraints to reduce the graph. \texttt{VertexReduction} is shown as Algorithm \ref{alg:VertexReduction}. First, we delete those vertices whose degree does not meet these requirements. Then, we can iterate over the neighbors of the deleted vertex and reduce their corresponding degrees by one. By doing so, some new vertices that do not satisfy the condition will appear. We recursively delete the vertices that do not satisfy the conditions until all the remaining vertices satisfy the conditions.

In Algorithm \ref{alg:VertexReduction}, a queue is used to store vertices that
should be removed. Since each vertex is pushed in and
popped from the queue at most once, the total processing time is $O(n)$. After that, if a vertex is removed, we need to
update the degrees for its neighbors. The total time cost is
$O(m)$. Therefore, the time complexity of Algorithm 4 is $O(n+m)$.

\stitle{Dichromatic Reduction (DR).} VR can cut the size of the graph, but it is not enough. We also propose a new reduction for each specific vertex.

\begin{lemma}[Pruning Rule in Unsigned Graph\cite{zhou2020enumerating}]\label{lem:Pruning Rule} 
In the undirected unsigned graph $G = (V,E)$, if $K$ is a $k$-plex and $|K| \geq q$ $(q\geq 2k-2)$, any other vertex $u$ which satisfies any of the following conditions is not in the $K$.

\ding{172} $u\in N_{G}(v)$ and $|N_{G}(u)\cap N_{G}(v)|<q-2k$

\ding{173} $u\in N_{G}^{2}(v)$ and $|N_{G}(u)\cap N_{G}(v)|<q-2k+2$

\end{lemma}
\begin{algorithm} [t!]
\footnotesize 
\caption{\texttt{DichromaticOnehop}$(nv,k,t)$}
\label{alg:Dichromatic-onehop}

\begin{algorithmic}[1]
	\State $g$ is the Dichromatic-network of $N_{G}\left(nv\right)$
	\For{$u \in N_{G}^{+}\left(nv\right)$}
	\State $L_{1}=L_{1} \cup \left \{ v \right \}$
	\EndFor
	\For{$u \in N_{G}^{-}\left(nv\right)$}
	\State $R_{1}=R_{1} \cup \left \{ v \right \}$
	\EndFor
	\While{$\exists v \in g, d_{g}^{+}(v)<t-2k$ or $ d_{g}^{+}(v)+d_{g}^{-}(v)<2t-2k$}
	\For{$u \in N_{g}^{+}(v)$}
	\State $d_{g}^{+}(u)-=1$
	\EndFor
	\For{$u \in N_{g}^{-}(v)$}
	\State $d_{g}^{-}(u)-=1$
	\EndFor
	\If{$v\in L_{1}$}
	\State $L_{1}\gets L_{1} \setminus \left\{v\right\}$
	\Else
	\State $R_{1}\gets R_{1} \setminus \left\{v\right\}$
	\EndIf
	\State update $g$
	\EndWhile
	\State \textbf{Return} \texttt{DichromaticTwohop}$(nv,L_{1},R_{1},k,t)$

\end{algorithmic}

\end{algorithm}
\begin{figure}[t!]\vspace*{0.2cm}
\centering
\includegraphics[height=5cm]{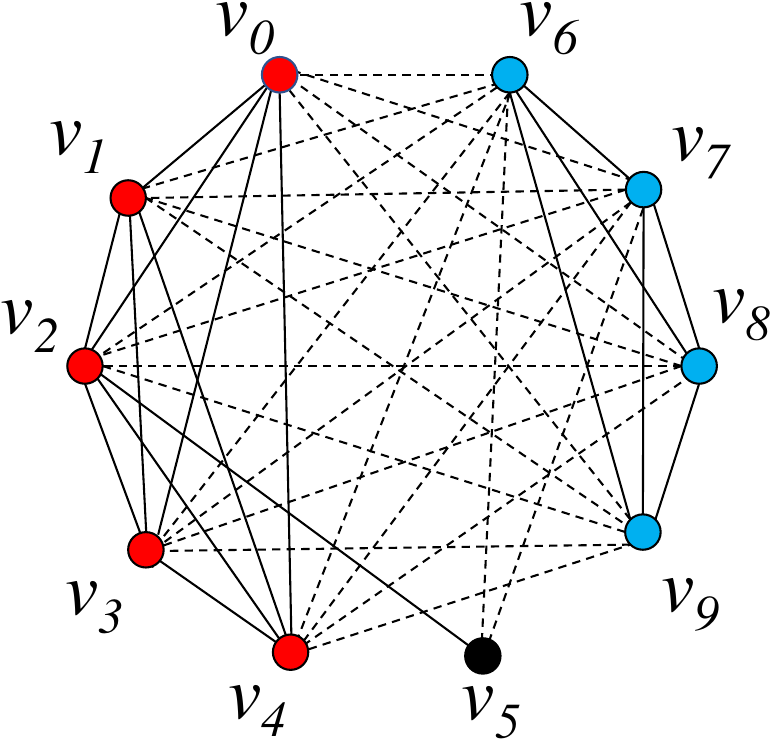}
\vspace*{-0.3cm} \caption{DichromaticOnehop example}
\label{fig:onehop} 
\end{figure}
%\begin{algorithm} [H]\label{alg:Dichromatic-onehop}
%	\caption{$Dichromatic-onehop(nv)$}
%
%	
%	\begin{algorithmic}[1]
%		
%		\For{$u \in N_{G}^{+}\left(nv\right)$}
%		\State $L=L \cup \left \{ v \right \}$
%		\EndFor
%		\For{$u \in N_{G}^{-}\left(nv\right)$}
%		\State $R=R \cup \left \{ v \right \}$
%		\EndFor
%		\State $qL=\emptyset$,$qR=\emptyset$
%		\For{$u \in L$}
%		\State $LPN[u]=N_{G}^{+}\left(u\right) \cap L$
%		\State $LNN[u]=N_{G}^{-}\left(u\right) \cap R$
%		\State $LPdegree[u]=LPN[u].size()$
%		\State $LNdegree[u]=LNN[u].size()$
%		\If{$LPdgredd[u] < T-2*K$ or $LPdegree[u]+LNdegree[u] < 2*T-2*K$}
%		\State $qL.push(u)$
%		\EndIf
%		\EndFor
%		\For{$u \in R$}
%		\State $RPN[u]=N_{G}^{+}\left(u\right) \cap R$
%		\State $RNN[u]=N_{G}^{-}\left(u\right) \cap L$
%		\State $RPdegree[u]=RPN[u].size()$
%		\State $RNdegree[u]=RNN[u].size()$
%		\If{$RPdegree[u] < T-2*K$ or $RPdegree[u]+RNdegree[u] < 2*T-2*K$}
%		\State $qR.push(u)$
%		\EndIf
%		\EndFor
%		\While{$qL \neq \emptyset$ or $qR \neq \emptyset$}
%		\If{$qL \neq \emptyset$}
%		\State $u=qL.front()$
%		\For{$v \in LPN[u]$}
%		\State $LPdegree[v]--$
%		\If{$LPdegree[v]<T-2*K$}
%		\State $qL.push(v)$
%		\EndIf
%		\EndFor
%		\For{$v \in LNN[u]$}
%		\State $RNdegree[v]--$
%		\If{$RNdegree[v]+RPdegree[v]<2*T-2*K$}
%		\State $qR.push(v)$
%		\EndIf
%		\EndFor
%		\State $qL.pop()$
%		\State $L=L-\left \{ v \right \}$
%		\EndIf
%		\If{$qR \neq \emptyset$}
%		\State same as $qL$
%		\EndIf
%		
%		\EndWhile
%		\Return $Dichromatic-twohop(L,R)$
%		
%		
%	\end{algorithmic}
%	
%\end{algorithm}

We apply the pruning rule in the Dichromatic-network\cite{yaocomputing}. For a vertex $v$, the ego network consists of $v$ and all its neighbor vertices, as well as all the edges between these vertices. For a vertex $v$, its Dichromatic-network is its Ego-network after removing all conflicting edges. The conflicting edges are negative edges between vertices of $N_{G}^{+}(u)$, negative edges between vertices of $N_{G}^{-}(u)$ and positive edges between a vertex of $N_{G}^{+}(u)$ and a vertex of $N_{G}^{-}(u)$.

We use Dichromatic-network instead of ego-network for pruning. Those conflicting edges are impossible to add into an antagonistic $k$-plex. If we use ego-network, it would be difficult to judge whether a vertex should be added to $C_{L}$ or $C_{R}$.
For example, if $v$ is the two-hop neighbor of $u$, $\exists w\in N_{G}^{+}(v) ,u\in N_{G}^{+}(w)$ and $\exists x\in N_{G}^{+}(v) ,u\in N_{G}^{-}(x)$, it is clear that $w$ is positive neighbor of $u$ and they should be in the same group. $x$ is negative neighbor of $u$, and they should be in different groups. However,  it is hard to say which group $v$ should be assigned. On the one hand, $v$ is a positive neighbor of $w$, so they should be in the same group. $u$ and $w$ are in the same group, so $u$ and $v$ should be in the same group. On the other hand,  $v$ is a positive neighbor of $x$, so they should be in the same group. $u$ and $x$ are in different groups, so $u$ and $v$ should be in different groups. In the ego-network, analyzing from different directions will lead to opposite conclusions about the relationship between $u$ and $v$. However, we can judge the vertex easily in Dichromatic-network. This is because there are no conflicting edges and no misunderstandings.

\begin{lemma} 
Finding an antagonistic $k$-plex in the Dichromatic-network is a necessary but insufficient condition for finding it on the original graph.

\end{lemma}

\begin{proof}
If we can find an antagonistic $k$-plex in Dichromatic-network, it does not mean that we can find this plex in the original graph due to conflicting edges. A conflicting edge can make a whole plex in Dichromatic-network unqualified. Fortunately, we find that if a vertices group cannot build a plex in Dichromatic-network, it cannot build a plex in ego-network. We can use this property to delete vertices on the Dichromatic-network and map them to the original graph.
\end{proof}

Next, we will show how to prune in the Dichromatic-network. First, we consider the one-hop neighbor. We have the following lemma:

\begin{theorem}[Pruning Rule for One-hop]\label{lem:Pruning Rule for one-hop} 
In the undirected signed graph $G = (V,E)$, if $C$ is a \mokplex, for a given $v\in C$, any other vertex $u$ which satisfies either of the following conditions is not in the $C$.

\ding{172} $u\in N_{G}^{+}(v)$ and $|N_{G}^{+}(u)\cap N_{G}^{+}(v)|<t-2k$

\ding{173} $u\in N_{G}^{-}(v)$ and $|N_{G}^{+}(u)\cap N_{G}^{-}(v)|<t-2k$

\ding{174} $u\in N_{G}(v)$ and $|N_{G}(u)\cap N_{G}(v)|<2t-2k$

\end{theorem}

\begin{proof}
According to Lemma \ref{lem:Pruning Rule}, each one-hop neighbor $v$ of a given $u$ should meet the requirements of the two small positive plexes and the whole antagonistic plex. For the small positive plex $C_{s}$ that $v$ belongs to, the number of positive neighbors of $v$ in $C_{s}$ is at least $|C_{s}|-2k$. For the whole antagonistic plex $C_{w}$ that $v$ belongs to, the number of neighbors of $v$ in $C_{w}$ is at least $|C_{w}|-2k$. Finally, according to Definition \ref{def:mokplex}, $C_{s}$ is at least $t$, $C_{w}$ is at least $2t$.
\end{proof}

\stitle{\underline{Algorithm Implementation.}} Reduction on one-hop neighbors is shown as Algorithm \ref{alg:Dichromatic-onehop}. First, we save all the positive and negative neighbors of $nv$ in $L$ and $R$, respectively (Lines 2--5). 
Then, we delete all unqualified vertices in $g$. However, when a vertex is deleted, it will affect the neighbor numbers of its neighbors. Some of its neighbors may become unqualified after deleting some vertices. At that time, we can add these vertices to the delete queue. Utill the delete queue is empty, we can stop, and all the remaining vertices in $L$ and $R$ are qualified one-hop neighbors.

\begin{example}
Figure \ref{fig:onehop} shows the example of Dichromatic-onehop reduction. We set $k=2,t=4$ and set $v_{0}$ as the current $nv$ in Algorithm \ref{alg:Dichromatic-onehop}. The other vertices in the graph are divided into two groups according to their adjacency to vertex $v_{0}$. The positive neighbours of $v_{0}$ are $v_{1},v_{2},v_{3},v_{4},v_{5}$. Negative neighbours are $v_{6},v_{7},v_{8},v_{9}$. However, the $|N_{G}^{+}(v_{5})\cap N_{G}^{+}(v_{0})|=3 < 2t-2k=4$. $v_{5}$ do not satisfy \ding{174} in lemma \ref{lem:Pruning Rule for one-hop}, so $v_{5}$ is removed from $L_{1}$. All the remaining vertices form the Dichromatic-onehop neighbors of $v_{0}$.
\end{example}

This process is similar to Algorithm \ref{alg:VertexReduction}. It can be seen as VR in the Dichromatic-network. However, Dichromatic-network is smaller than $G$. There are at most $\Delta$ vertices and $\Delta^{2}$ edges in Dichromatic-network. So the time complexity of Algorithm \ref{alg:Dichromatic-onehop} is $O(\Delta+\Delta^{2})$, when $\Delta$ is the maximum degree of $v\in G$. 
%\textcolor{red}{[explain $\Delta$]}

\begin{figure}[t!]\vspace*{-0.3cm}
\centering
\includegraphics[height=5cm]{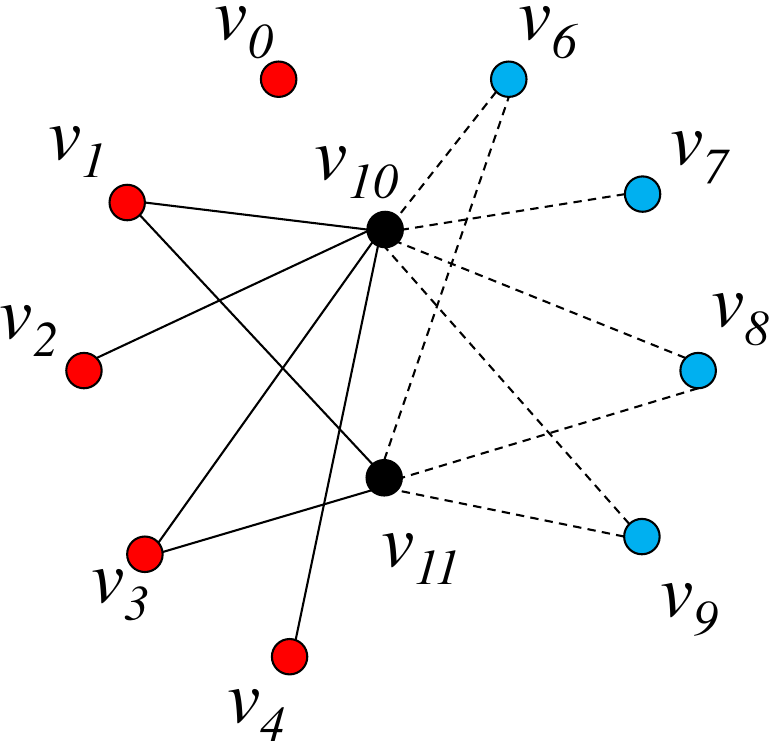}
\vspace*{-0.3cm} \caption{DichromaticTwohop example}
\label{fig:twohop} \vspace*{-0cm}
\end{figure}

\begin{algorithm} [t!]
\footnotesize
\caption{\texttt{DichromaticTwohop}$(nv,L_{1},R_{1},k,t)$}
\label{alg:Dichromatic-twohop}

\begin{algorithmic}[1]
	\State $Ln=\emptyset$,$Rn=\emptyset$
	\For{$u \in L_{1}$}
	\For{$v \in N_{G}^{+}\left(u\right)$}
	\State $L_{2}=L_{2}\cup \left \{ v \right \}$
	\EndFor
	\For{$v \in N_{G}^{-}\left(u\right)$}
	\State $R_{2}=R_{2}\cup \left \{ v \right \}$
	\EndFor
	\EndFor
	\For{$u \in R_{1}$}
	\For{$v \in N_{G}^{+}\left(u\right)$}
	\State $R_{2}=R_{2}\cup \left \{ v \right \}$
	\EndFor
	\For{$v \in N_{G}^{-}\left(u\right)$}
	\State $L_{2}=L_{2}\cup \left \{ v \right \}$
	\EndFor
	\EndFor
	\For{$v \in L_{2} \setminus N_{G}\left(nv\right)$}
	\State $a=|N_{G}^{+}\left(u\right) \cap L_{1}|$
	\State $b=|N_{G}^{-}\left(u\right) \cap R_{1}|$
	\If{$a\geq t-2*k+2$ and $a+b \geq 2*t-2*k+2$}
	\State $Ln=Ln\cup \left \{ v \right \}$
	\EndIf
	\EndFor
	\For{$v \in R_{2} \setminus N_{G}\left(nv\right)$}
	\State $a=|N_{G}^{+}\left(u\right) \cap R_{1}|$
	\State $b=|N_{G}^{-}\left(u\right) \cap L_{1}|$
	\If{$a\geq t-2*k+2$ and $a+b \geq 2*t-2*k+2$}
	\State $Rn=Rn\cup \left \{ v \right \}$
	\EndIf
	\EndFor
	\State \textbf{Return} $Ln=Ln\cup L_{1},Rn=Rn\cup R_{1}$
	
\end{algorithmic}

\end{algorithm}

\begin{theorem}[Pruning Rule for Two-hop]\label{lem:Pruning Rule for two-hop} 
In the undirected signed graph $G = (V,E)$, if $C$ is a \mokplex, all qualified one-hop neighbors are in $L_{1}\cup R_{1}$, $w$ is the neighbor of a vertex in $L_{1}\cup R_{1}$. For all the following cases, $w$ is not in $C$, if $a<t-2k+2$ or $a+b<2t-2k+2$.

\ding{172} $v\in L_{1}$, $w\in N_{G}^{+}\left(v\right)$. let $a$ be the number of $w$'s positive neighbors in $L_{1}$, $b$ be the number of $w$'s negative neighbors in $R_{1}$.

\ding{173} $v\in L_{1}$, $w\in N_{G}^{-}\left(v\right)$. let $a$ be the number of $w$'s positive neighbors in $R_{1}$, $b$ be the number of $w$'s negative neighbors in $L_{1}$.

\ding{174} $v\in R_{1}$, $w\in N_{G}^{+}\left(v\right)$. let $a$ be the number of $w$'s positive neighbors in $R_{1}$, $b$ be the number of $w$'s negative neighbors in $L_{1}$.

\ding{175}$v\in R_{1}$, $w\in N_{G}^{-}\left(v\right)$. let $a$ be the number of $w$'s positive neighbors in $L_{1}$, $b$ be the number of $w$'s negative neighbors in $R_{1}$.

\end{theorem}

\begin{proof}
First, we have identified all Dichromatic-onehop neighbors. Two-hop neighbors are neighbors of one-hop neighbors. After Algorithm \ref{alg:Dichromatic-onehop} is completed, we have $L_{1}$ as the qualified one-hop positive neighbors and $R_{1}$ as the qualified one-hop negative neighbors. We can divide Dichromatic-twohop neighbors into four parts according to the adjacency with the Dichromatic-onehop neighbors. They are positive neighbors of vertices in $L_{1}$, negative neighbors of vertices in $L_{1}$, positive neighbors of vertices in $R_{1}$, and negative neighbors of vertices in $R_{1}$. According to Lemma \ref{lem:Pruning Rule}, each Dichromatic-twohop neighbor $v$ of a given $u$ should also meet the requirements of two small positive plexes and the whole antagonistic plex. For the small positive plex $C_{s}$ that $v$ belongs to, $v$'s positive neighbors in $C_{s}$ is at least $|C_{s}|-2k+2$. For whole antagonistic plex $C_{w}$ that $v$ belongs to, $v$'s neighbors in $C_{w}$ is at least $|C_{w}|-2k+2$.
\end{proof}

\stitle{\underline{Algorithm Implementation.}} Reduction on two-hop neighbors is shown as Algorithm \ref{alg:Dichromatic-twohop}. We use $Ln$ and $Rn$ to store the final candidate sets of the given vertex $nv$. We put the positive neighbors of vertices in $L_{1}$ and the negative neighbors of vertices in $R_{1}$ into $L_{2}$ as candidate sets for 2-hop vertices in $L_{n}$. Similarly, we put the positive neighbors of the vertices of $R_{1}$ and the negative neighbors of the vertices of $L_{1}$ into $R_{2}$ as the candidate set of 2-hop vertices in $R_{n}$ (Line 2--11). Then, we traverse $L_{2}$ and $R_{2}$ respectively and put the vertices meeting the above conditions into $L_{n}$ and $R_{n}$ respectively (Line 12--21). Then, we union the two-hop candidate $Ln$ and $Rn$ with one-hop candidate $L_{1}$ and $R_{1}$, respectively (Line 22). Finally, we get $Ln$ and $Rn$ as the candidate vertices sets for the given vertex $nv$.

\begin{example}
Figure \ref{fig:twohop} shows an example of Dichromatic-twohop reduction. We set $k=2,t=4$ and set $v_{0}$ as the current $nv$ in Algorithm \ref{alg:Dichromatic-twohop}. $L_{1}$ and $R_{1}$ are the Dichromatic-onehop neighbor sets of $nv$. This figure omits the lines among $v_{0}$ and Dichromatic-onehop neighbor of $v_{0}$. $v_{10}$ is the two-hop neighbor of $v_{0}$. It has four positive neighbors in $L_{1}$ and negative neighbors in $R_{1}$, which satisfy the requirements of Dichromatic-twohop. However, $|N_{G}^{+}\left(v_{11}\right) \cap L_{1}|+|N_{G}^{-}\left(v_{11}\right) \cap R_{1}|=5<2t-2k+2=6$. Therefore, $v_{11}$ is not Dichromatic-twohop neighbor of $v_{0}$.
\end{example}

In DR, we need to iterate over all the neighbors of the two-hop neighbors of $nv$. So the time complexity of Algorithm \ref{alg:Dichromatic-twohop} is $O(\Delta^{3})$. As the DR is called for each vertex in $G$, the total time complexity of DR is $O(\Delta^{3}n)$.

%\begin{algorithm} [H]
%	\caption{$Dichromatic-twohop(L,R,k,t)$}
%	\label{alg:Dichromatic-twohop}
%	
%	\begin{algorithmic}[1]
%		\State $Ln=\emptyset$,$Rn=\emptyset$
%		\For{$u \in L$}
%		\For{$v \in N_{G}^{+}\left(u\right)$}
%		\State $a=|N_{G}^{+}\left(u\right) \cap L|$
%		\State $b=|N_{G}^{-}\left(u\right) \cap R|$
%		\If{$a\geq t-2*k$ or $a+b \geq 2*t-2*k$}
%		\State $Ln=Ln\cup \left \{ v \right \}$
%		\EndIf
%		\EndFor
%		\For{$v \in N_{G}^{-}\left(u\right)$}
%		\State $a=|N_{G}^{+}\left(u\right) \cap R|$
%		\State $b=|N_{G}^{-}\left(u\right) \cap L|$
%		\If{$a\geq t-2*k$ or $a+b \geq 2*t-2*k$}
%		\State $Rn=Rn\cup \left \{ v \right \}$
%		\EndIf
%		\EndFor
%		\EndFor
%		\For{$u \in R$}
%		\For{$v \in N_{G}^{+}\left(u\right)$}
%		\State $a=|N_{G}^{+}\left(u\right) \cap R|$
%		\State $b=|N_{G}^{-}\left(u\right) \cap L|$
%		\If{$a\geq t-2*k$ or $a+b \geq 2*t-2*k$}
%		\State $Rn=Rn\cup \left \{ v \right \}$
%		\EndIf
%		\EndFor
%		\For{$v \in N_{G}^{-}\left(u\right)$}
%		\State $a=|N_{G}^{+}\left(u\right) \cap L|$
%		\State $b=|N_{G}^{-}\left(u\right) \cap R|$
%		\If{$a\geq t-2*k$ or $a+b \geq 2*t-2*k$}
%		\State $Ln=Ln\cup \left \{ v \right \}$
%		\EndIf
%		\EndFor
%		\EndFor
%		\Return $Ln=Ln\cup L,Rn=Rn\cup R$
%		
%	\end{algorithmic}
%	
%\end{algorithm}
\stitle{\underline{Algorithm Implementation Summary.}} We summarize the preprocessing algorithm and show it as Algorithm \ref{alg:SAPE}. In the first step, we use VR (Line 1). Then, for each vertex we use Dichromatic Reduction to reduce the number of the vertices in the candidate set (Line 5). After that, $P_{L}$, $P_{R}$, $Q_{L}$ and $Q_{R}$ is computed (Line 6--9) and \texttt{SAPEUTIL} is called (Line 10).

\begin{algorithm}[t!] 
\footnotesize
\caption{\texttt{SAPE}$(G=(V,E^{+},E^{-}),k,t)$}
\label{alg:SAPE} 
\hspace*{\algorithmicindent} \textbf{Input: }a signed graph $G$,$k$,$t$ \\
\hspace*{\algorithmicindent} \textbf{Output: }All maximal antagonistic $k$-plex
\begin{algorithmic}[1]
	\State \texttt{VertexReduction}$(G,k,t)$
	\State $Flag \gets  true$ 
	\For{$v_{i}\in \left \{ v_{0},v_{1},...,v_{n-1}\right \}$}
	\State $C_{L}\gets  \left \{v_{i} \right \},C_{R}\gets  \emptyset$
	\State $Ln,Rn=$\texttt{DichromaticOnehop}$(v_{i},k,t)$
	\State $P_{L}\gets Ln \cap \left \{ v_{i+1},...,v_{n-1}\right \}$
	\State $P_{R}\gets Rn \cap \left \{ v_{i+1},...,v_{n-1}\right \}$
	\State $Q_{L}\gets Ln \cap \left \{ v_{0},...,v_{i-1}\right \}$
	\State $Q_{R}\gets Rn \cap \left \{ v_{0},...,v_{i-1}\right \}$
	\State \texttt{SAPEUTIL}$\left ( C_{L},C_{R},P_{L},P_{R},Q_{L},Q_{R}\right )$
	\EndFor

\end{algorithmic}

\end{algorithm}

It is worth noting that although a variety of optimization methods are incorporated in Algorithm \ref{alg:SAPE} compared to Algorithm \ref{alg:BAPE}, Algorithm \ref{alg:SAPE} still uses the structure of set enumeration and its worst time complexity does not change. However, it can significantly reduce the number of iterations of the set enumeration. Algorithm \ref{alg:SAPE} will be more efficient than Algorithm \ref{alg:BAPE} in actual computation.

\section{EXPERIMENTS}

\begin{figure*}[t!] 
\centering
\subfigcapskip=-9pt
\subfigure[{\scriptsize Slashdot, $k=2$}]{
	\includegraphics[width=4.12cm]{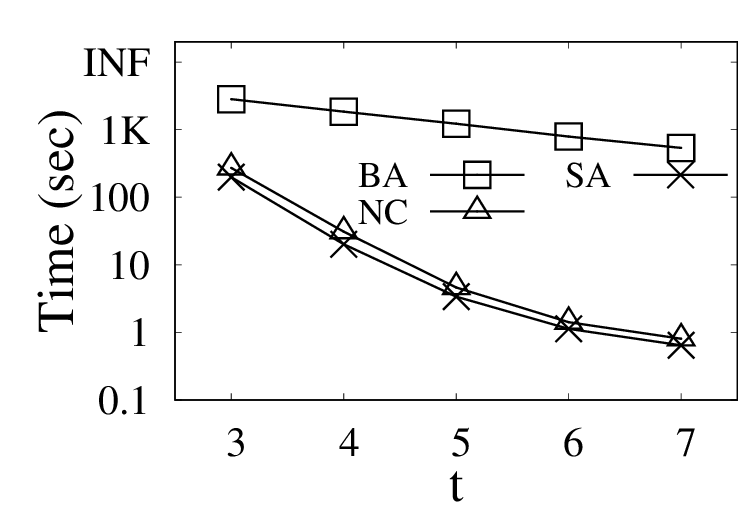}
}
\subfigure[{\scriptsize Epinions, $k=2$}]{
	\includegraphics[width=4.12cm]{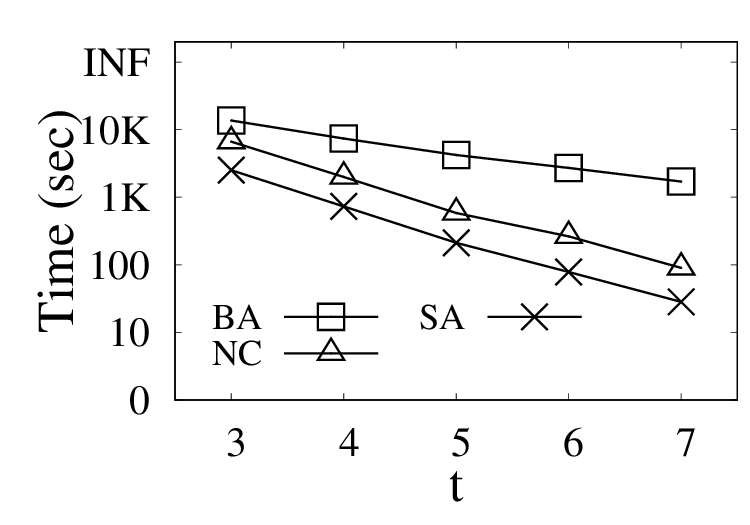}
}
\subfigure[{\scriptsize Super, $k=2$}]{
	\includegraphics[width=4.12cm]{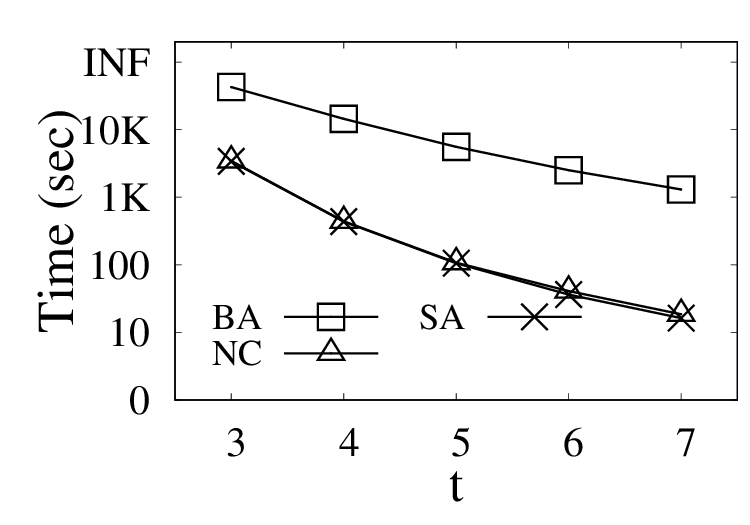}
}
\subfigure[{\scriptsize WiKi, $k=2$}]{
	\includegraphics[width=4.12cm]{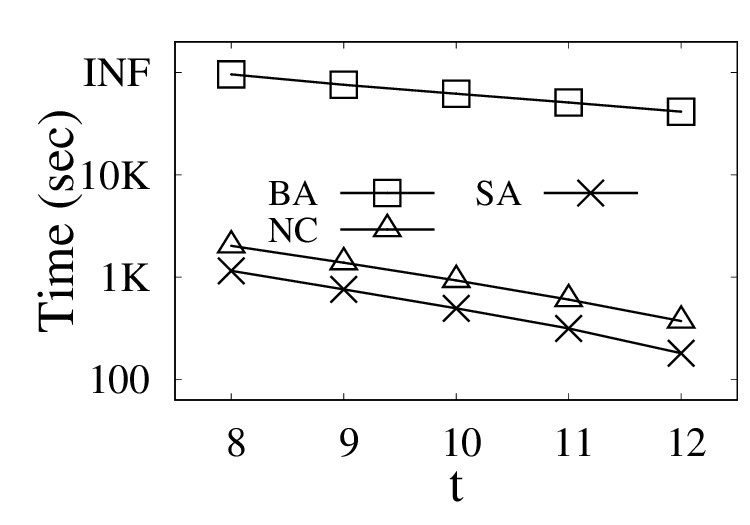}
}\vspace*{-0.3cm}
\subfigure[{\scriptsize Slashdot, $k=3$}]{
	\includegraphics[width=4.12cm]{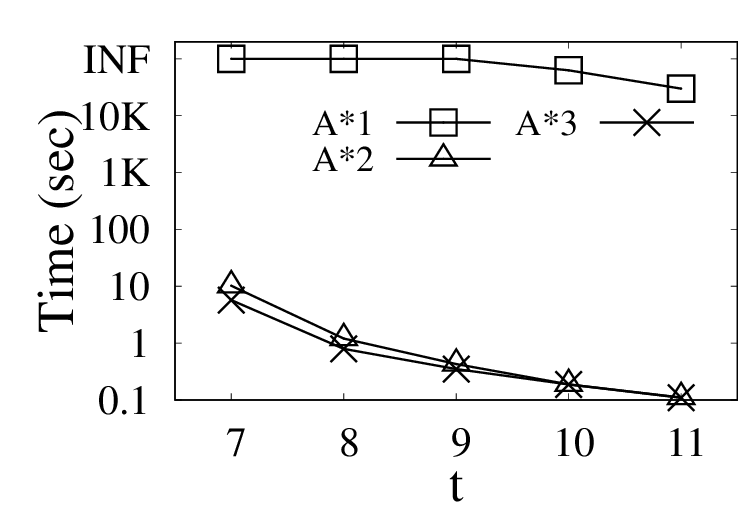}
}
\subfigure[{\scriptsize Epinions, $k=3$}]{
	\includegraphics[width=4.12cm]{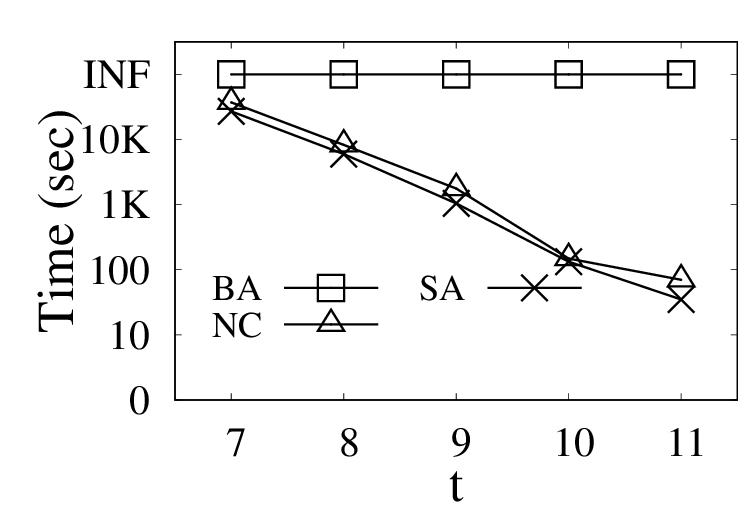}
}
\subfigure[{\scriptsize Super, $k=3$}]{
	\includegraphics[width=4.12cm]{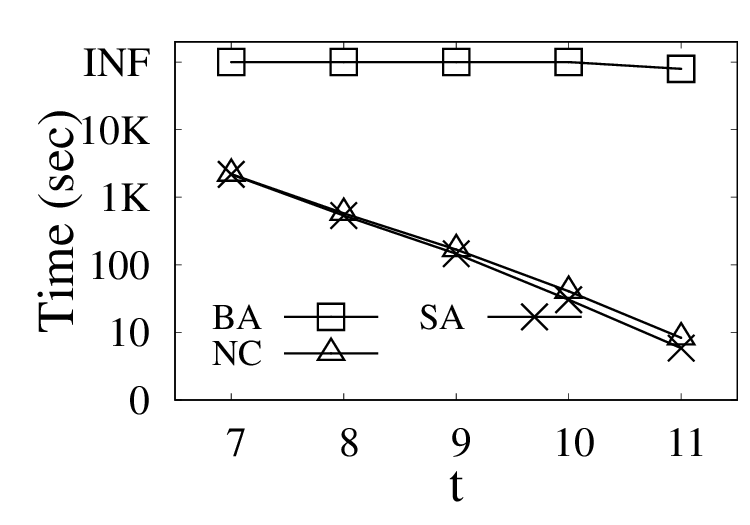}
}
\subfigure[{\scriptsize WiKi, $k=3$}]{
	\includegraphics[width=4.12cm]{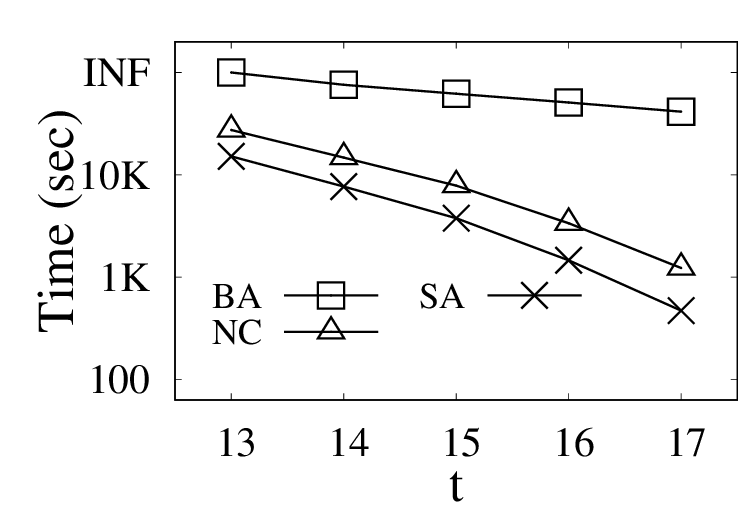}
}
\vspace*{-0.3cm}
\caption{Running time of different algorithms varying $t$,$k$}
\label{fig:Running time of different algorithms}
\end{figure*}

In this section, we present our experimental results. All the experiments are performed on a machine with Intel(R) Xeon(R) Gold 5218R CPU @ 2.10GHz and 96GB RAM and Ubuntu system. All algorithms are implemented in C++, using g++ complier with -O3. The time cost is measured as the amount of wall-clock time elapsed during the program’s execution. If an algorithm cannot finish in 12 hours, we denote the processing time as INF. We evaluate our algorithms in some real and synthetic signed networks.

%\textbf{Algorithms.} We compare two algorithms: BAPE (SA), SAPE (SA) and SANC(NC). BAPE is the baseline solution shown in Section 3. SAPE is the algorithm with the enumeration optimization shown in Section 4. SANC is the algorithm with all the enumeration optimization but color bound pruning. Note that the preprocessing optimization strategies can be also used in BAPE, thus, we apply them for both algorithms for fairness.

\stitle{Algorithms.} We evaluate the performance of the following methods:

\begin{itemize}
% \vspace{-0.4em}
\item \texttt{Baseline.} It is the basic solution shown in Section 3. The baseline execution is too slow and cannot be completed within 12 hours. We use BA instead of baseline.
\item \texttt{BAPE.} Apply VR to baseline. For simplicity, we denote it by BA in the following figures.
\item \texttt{SANC.} It is the algorithm with all the optimization but color-bound pruning. We denote it by NC in the following figures.
\item \texttt{SAPE.} It is the algorithm with the enumeration optimization shown in Section 4. We denote it by SA in the following figures.

\end{itemize}

Note that the preprocessing optimization strategies VR can also be used in \texttt{BAPE}. Thus, we apply them to all three algorithms for fairness.

\stitle{Datasets.} We use four datasets downloaded from http://snap.stanford.edu to evaluate our algorithms. Slashdot and Epinions are real-world signed networks. Super and WiKi are not signed graphs, but we use the method in \cite{li2018efficient} to transform them into signed networks. The details of each dataset are shown in Table \ref{table:datasets}.

\begin{table}[t!] \vspace*{-0.4cm}
%\small
%\scriptsize
\centering
\caption{Datasets} \label{table:datasets}
\vspace*{-0.2cm}
\setlength{\tabcolsep}{1mm}{
	\begin{tabular}{c|c|c|c|c|c}
		\hline
		Dataset & $|V|$ &  $| E^{+}|$& $| E^{-}|$ &$max(d_{G}^{+})$&$max(d_{G}^{-})$ \\ \hline
		Slashdot	&77,357				&396,378		&120,197&2,507&598\\
		Epinions  			&131,828			&717,667	&123,705	&3,334&1,590\\
		
		Super  		&567,301			&82,547	&632,023&2,598& 11,696\\
		WiKi  		&1,140,149			&450,467	&2,337,500 &17,092&124,859\\
		
		\hline
	\end{tabular}
} 	
\end{table}

\subsection{Efficiency when varying $k$ and $t$} 

In this experiment, we evaluate the efficiency of three algorithms when varying $t$ and $k$. The results are shown in Figure \ref{fig:Running time of different algorithms}.

For each data set, we test five different $t$-values corresponding to $k=2$ and $k=3$. As shown in Figure \ref{fig:Running time of different algorithms}. Among all the test cases, \texttt{BAPE} is the least efficient of the three algorithms. The other two algorithms are considerably more efficient than \texttt{BAPE}. For example, when $t=3$ and $k=3$, \texttt{BAPE} cannot finish computing Slashdot in 12h, while \texttt{BAPE} and \texttt{SANC} can complete within minutes. This is mainly because \texttt{BAPE} computes too many hopeless search branches, while the other two algorithms use our proposed enumeration optimization. These optimizations help the algorithm estimate in advance whether the current search branch has any hope of finding \mokplex in the future. If the current search branch is no longer possible to find the qualified \mokplex, it is worthless to continue the computation in this time. It can be skipped without affecting the accuracy of the final result. Comparing \texttt{SAPE} and \texttt{SANC}, the only difference is the use of color-bound pruning. Calculating color bound increases the overhead of the algorithm, but it can further reduce the number of search branches and iterations at the same time. The efficiency of these two algorithms is relatively close, but as shown in the Figure \ref{fig:Running time of different algorithms}, the overall efficiency of \texttt{SAPE} is still improved by using color bound. 
%For example, when $t$=5 and $k=2$, SAPE can finish in 200 seconds while SANC needs 250 seconds. 
In addition, it can be seen from the figure that the efficiency of all algorithms increases either when the value of $k$ becomes smaller or the value of t becomes larger. This is due to the increased pruning ability of preprocessing optimization at this time. Therefore, the number of vertices that can enter the set enumeration phase is reduced significantly. This reduces the search space of the set enumeration.

%\textbf{Exp-2: Evaluation of VR}
\subsection{Evaluation of Vertex Reduction}

\begin{figure}[t!] \vspace*{-0.5cm}
\centering
\subfigcapskip=-9pt
\subfigure[{\scriptsize Slashdot}]{
	\includegraphics[width=4.12cm]{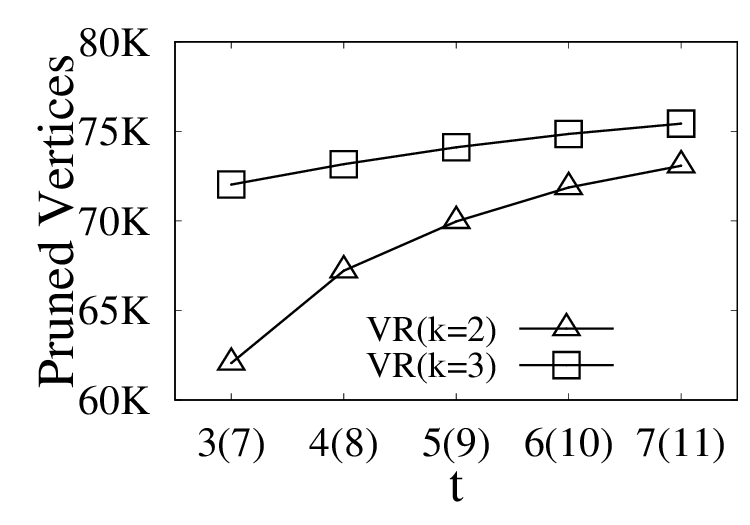}
}
\subfigure[{\scriptsize Epinions}]{
	\includegraphics[width=4.12cm]{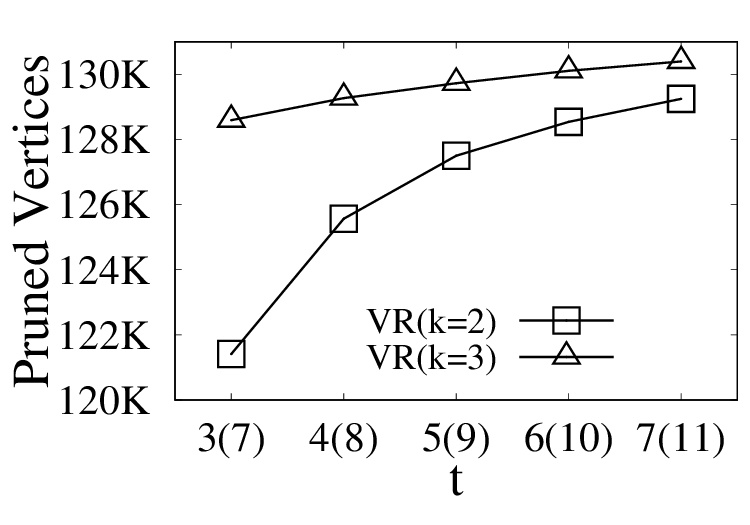}
}\vspace*{-0.4cm}
\subfigure[{\scriptsize Super}]{
	\includegraphics[width=4.12cm]{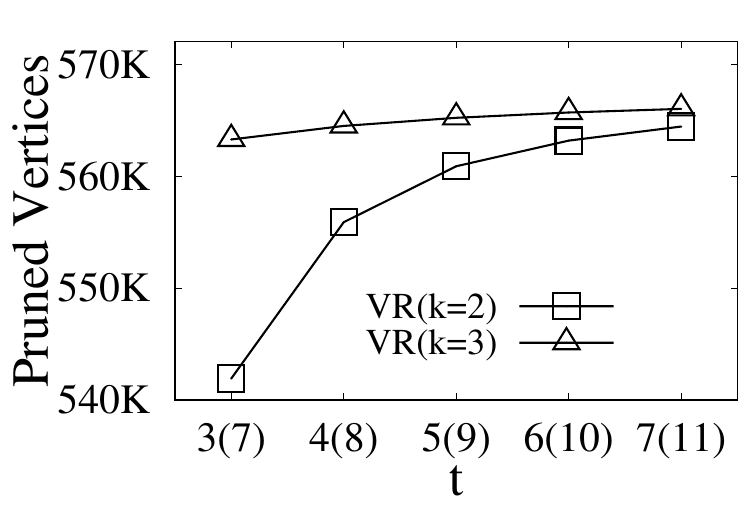}
}
\subfigure[{\scriptsize WiKi}]{
	\includegraphics[width=4.12cm]{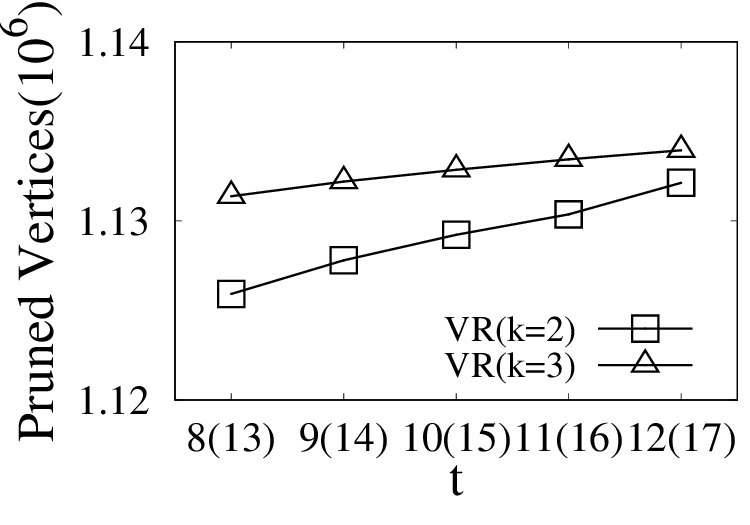}
}
\vspace*{-0.5cm}
\caption{Pruned vertices by VR}
\label{fig:Pruned vertices by VertexReduction} 
\end{figure}

\begin{figure}[t!] \vspace*{-0cm}
\centering
\subfigcapskip=-9pt
\subfigure[{\scriptsize Slashdot}]{
	\includegraphics[width=4.12cm]{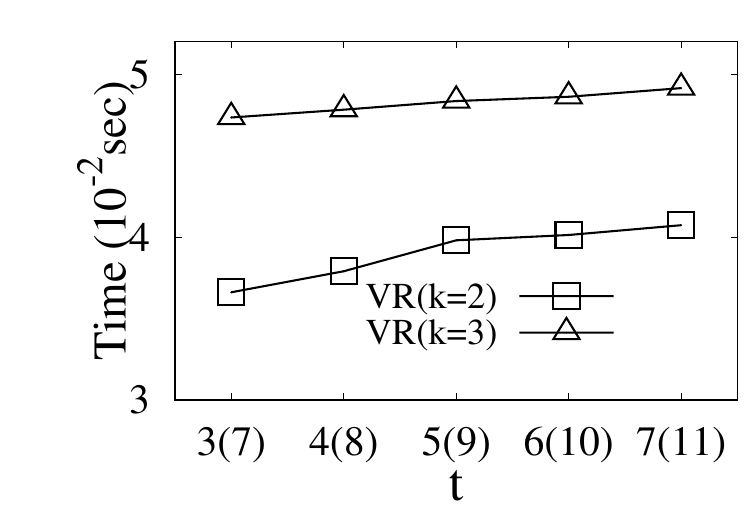}
}
\subfigure[{\scriptsize Epinions}]{
	\includegraphics[width=4.12cm]{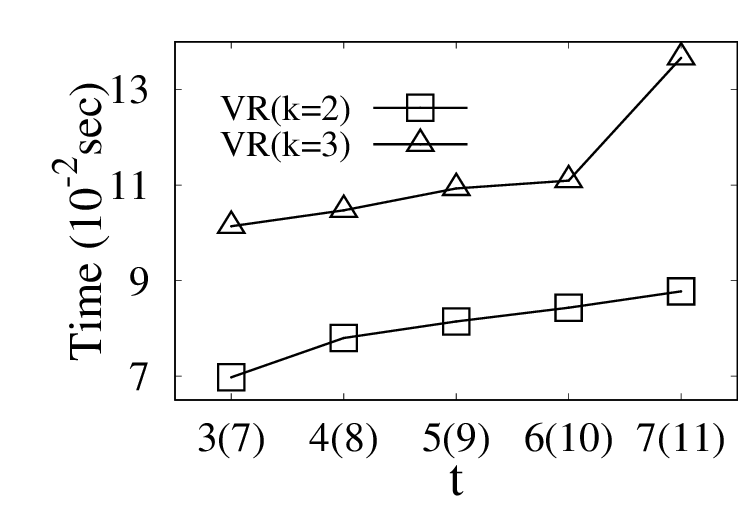}
}\vspace*{-0.4cm}
\subfigure[{\scriptsize Super}]{
	\includegraphics[width=4.12cm]{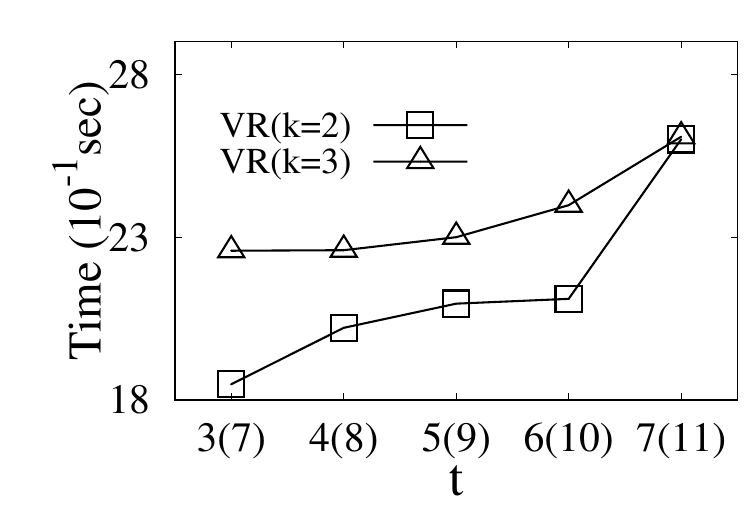}
}
\subfigure[{\scriptsize WiKi}]{
	\includegraphics[width=4.12cm]{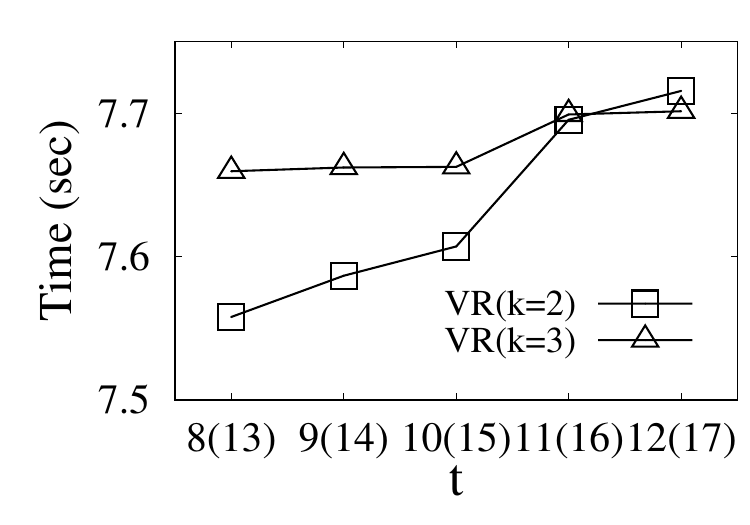}
}
\vspace*{-0.5cm}
\caption{Running time of VR}
\label{fig:Running time of VertexReduction} 
\end{figure}

In this experiment, we evaluate the effectiveness and efficiency of the
VR strategy. Results are shown in Figure \ref{fig:Pruned vertices by VertexReduction} and \ref{fig:Running time of VertexReduction}. In these figures, given the dataset and value of $k$, the value of $t$ corresponds to its setting in Figure \ref{fig:Running time of different algorithms}. For example, the corresponding $t$ of Slashdot $k=2$ are the value in front of the abscissa axis $\left\{3, 4, 5, 6, 7\right\}$, and the corresponding $t$ of Slashdot $k=3$ are the value behind the abscissa axis $\left\{7, 8, 9, 10, 11\right\}$. The same is true for other datasets. All subsequent experimental graphs also obey this setting. Figure \ref{fig:Pruned vertices by VertexReduction} shows the ability of VR strategy to delete vertices. For example, more than 90 percent of vertices can be pruned in Slashdot and Epinions. As the values of $k$ and $t$ increase, more and more vertices in the process of VR can be removed.

Figure \ref{fig:Running time of VertexReduction} shows the running time of VR. VR can remove millions of hopeless vertices in a very short period of time. As the value of $k$ decreases and the value of $t$ increases, more vertices need to be removed, and the running time of VR increases slightly, but the change is minimal.

\subsection{Evaluation of Dichromatic Reduction}

\begin{figure}[t!] \vspace*{-0cm}
\centering
\subfigcapskip=-9pt
\subfigure[{\scriptsize Slashdot}]{
	\includegraphics[width=4.12cm]{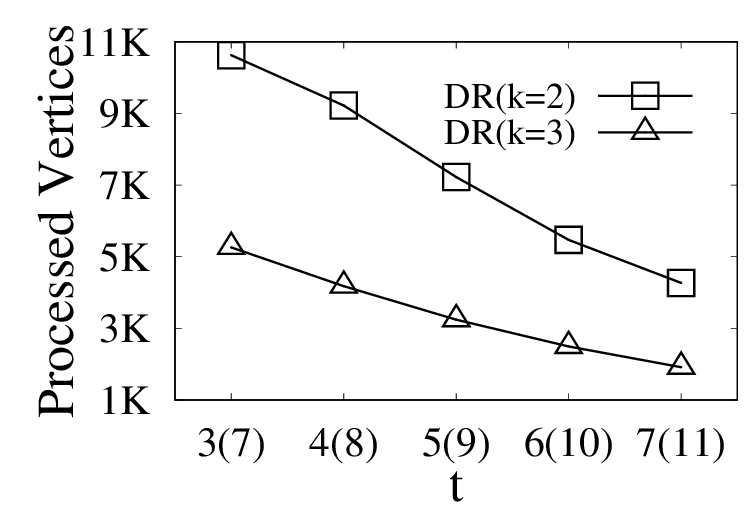}
}
\subfigure[{\scriptsize Epinions}]{
	\includegraphics[width=4.12cm]{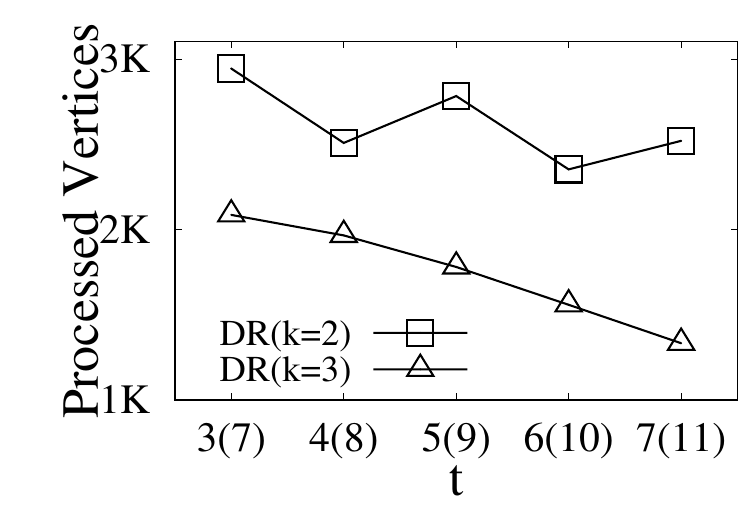}
}\vspace*{-0.4cm}
\subfigure[{\scriptsize Super}]{
	\includegraphics[width=4.12cm]{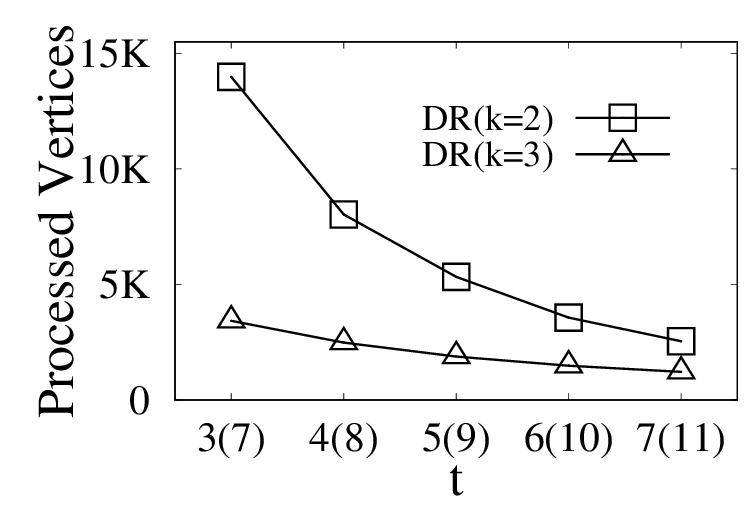}
}
\subfigure[{\scriptsize WiKi}]{
	\includegraphics[width=4.12cm]{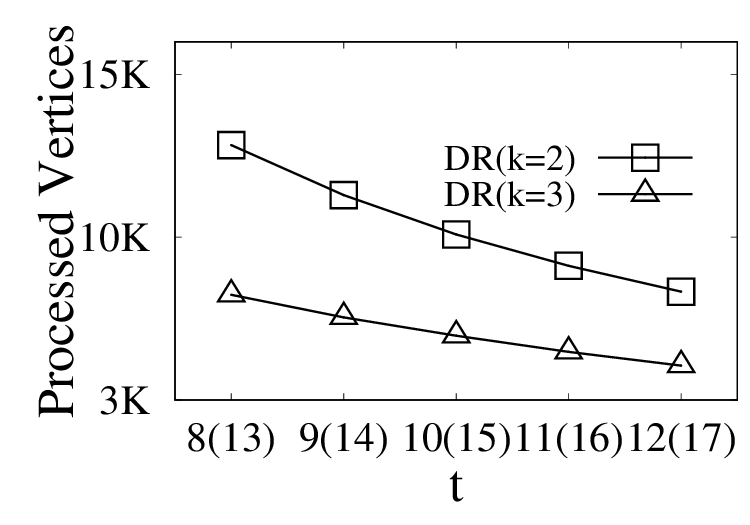}
}
\vspace*{-0.5cm}
\caption{Pruned vertices by DR}
\label{fig:Pruned vertices by DichromaticReduction} \vspace*{0.1cm}
\end{figure}

In this experiment, we evaluate the effectiveness and efficiency of the
DR strategies. Figure \ref{fig:Pruned vertices by DichromaticReduction} shows the number of processed vertices after using the DR strategy. After using the DR strategy in Algorithm \ref{alg:SAPE}, the sizes of the four sets $P_{L}$, $P_{R}$, $Q_{L}$ and $Q_{R}$ corresponding to each vertex are reduced. The candidate sets of many vertices no longer satisfy the expansion condition. Therefore, these vertices can be skipped during set enumeration. It can be seen from Figure \ref{fig:Pruned vertices by DichromaticReduction} that the DR strategy can further prune the remaining vertices after using the VR strategy on a large scale. With the increase of $k$ and $t$, the number of vertices pruned by DR tends to decrease, which is mainly due to the enhanced pruning ability of VR strategy.
%Figure \ref{fig:Pruned vertices by DichromaticReduction} shows the average of the ratio of the number of vertices remaining in the candidate set in each iteration after the DR process to the size of the original candidate set. Not applicable A, the size of the candidate set for each iteration of Algorithm \ref{alg:SAPE} is the number of two-hup neighbors of the enumerated vertices in this round. DR can further reduce the size of the candidate set before the set enumeration. In most experiments, on average, less than three percent of the vertices per iteration remained after the DR process.

\begin{figure}[t!] \vspace*{-0cm}
\centering
\subfigcapskip=-9pt
\subfigure[{\scriptsize Slashdot}]{
	\includegraphics[width=4.12cm]{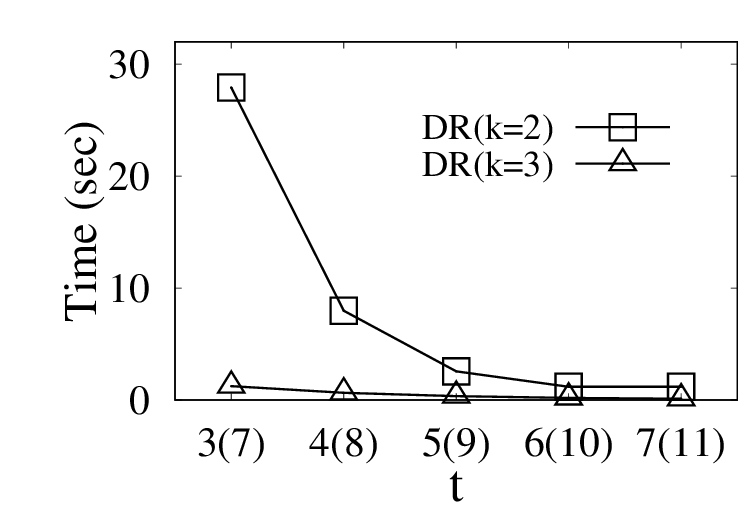}
}
\subfigure[{\scriptsize Epinions}]{
	\includegraphics[width=4.12cm]{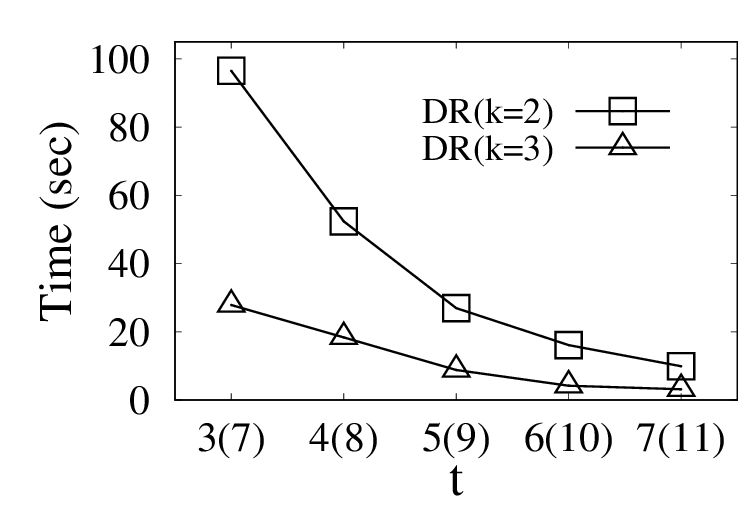}
}\vspace*{-0.4cm}
\subfigure[{\scriptsize Super}]{
	\includegraphics[width=4.12cm]{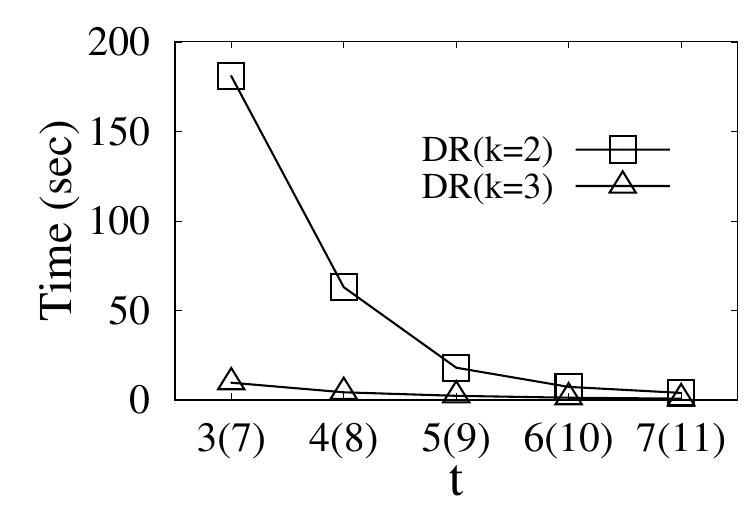}
}
\subfigure[{\scriptsize WiKi}]{
	\includegraphics[width=4.12cm]{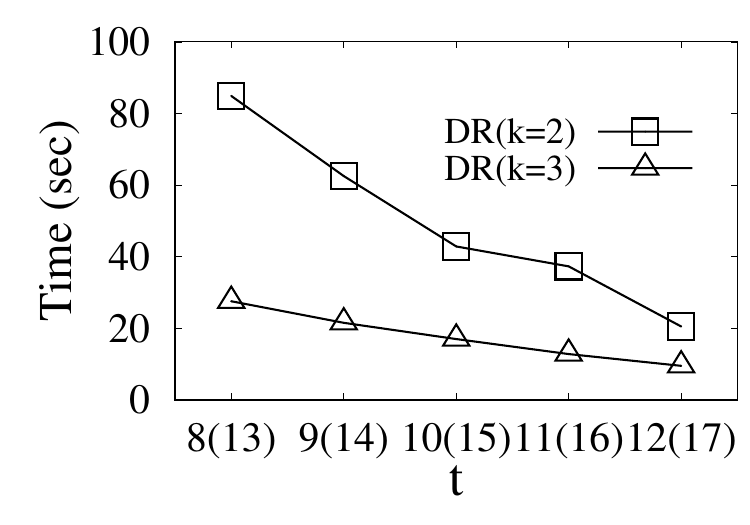}
}
\vspace*{-0.5cm}
\caption{Running time of DR}
\label{fig:Running time of DichromaticReduction}
\end{figure}

Figure \ref{fig:Running time of DichromaticReduction} shows the time of DR. As $k$ and $t$ increase, the running time of DR also decreases. After VR, the number of vertices becomes smaller and each vertex has fewer two-hop neighbors. It reduces the search space of DR.

\subsection{Scalability testing}

\begin{figure}[t!] \vspace*{-0cm}
\centering
\subfigcapskip=-9pt
\subfigure[{\scriptsize WiKi(vary e)}]{
	\includegraphics[width=4.12cm]{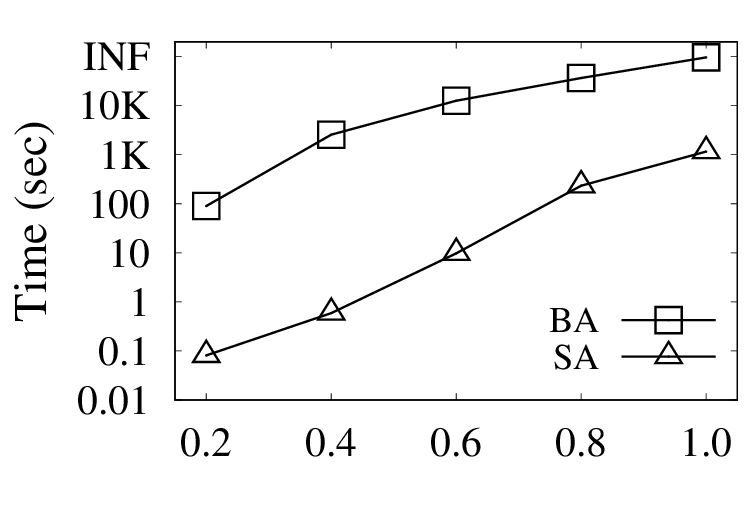}
}
\subfigure[{\scriptsize WiKi(vary n)}]{
	\includegraphics[width=4.12cm]{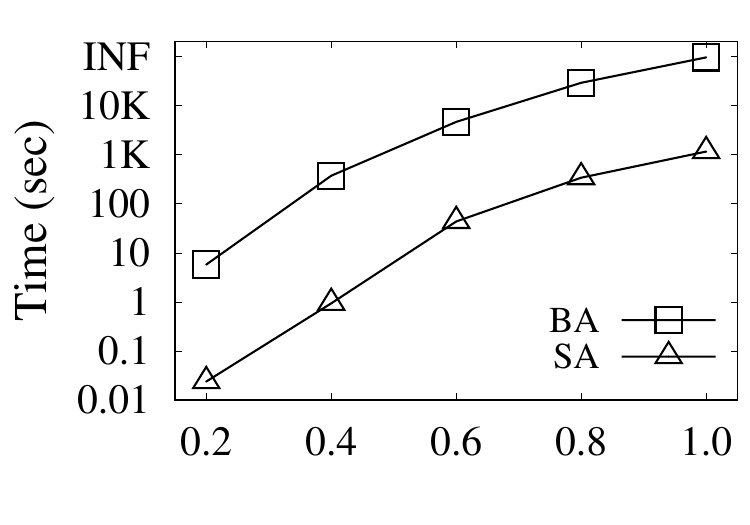}
}
\vspace*{-0.5cm}
\caption{Scalability of \texttt{BAPE} and \texttt{SAPE}}
\label{fig:Scalability of BAPE and SAPE} 
\end{figure}

In this experiment, we test the scalability of \texttt{BAPE} and \texttt{SAPE} on the large dataset WiKi by varying their vertices and edges from 20\% to 100\%. We set $t=8$ and $k=2$ as default values. Figure \ref{fig:Scalability of BAPE and SAPE} shows the results of the basic algorithm and our modified algorithms.

As shown in Figure \ref{fig:Scalability of BAPE and SAPE}, the running times of both algorithms increase as graph increases, but \texttt{SAPE} outperforms \texttt{BAPE} in all cases. e.g., when we sample 40\% of the vertices, the running times of \texttt{SAPE} and \texttt{BAPE} are 0.9 and 372 seconds, respectively, while when 80\% of the vertices are sampled, their runtimes are 232 and 36,458 seconds, respectively. This experiment shows that our modified algorithms have good scalability in practice.

%\begin{figure}[t!] \vspace*{-0.35cm}
%	\centering
%	\subfigure[{\scriptsize SN1(Vary $t$)}]{
%		\includegraphics[width=4cm]{figure/Exp6SN1}
%	}
%	\subfigure[{\scriptsize SN2(Vary $t$)}]{
%		\includegraphics[width=4cm]{figure/Exp6SN2}
%	}
%	\subfigure[{\scriptsize SN3(Vary $t$)}]{
%		\includegraphics[width=4cm]{figure/Exp6SN3}
%	}
%	\subfigure[{\scriptsize SN4(Vary $t$)}]{
%		\includegraphics[width=4cm]{figure/Exp6SN4}
%	}
%	\vspace*{-0.25cm}
%	\caption{Running time of different algorithms in synthetic datasets}
%	\label{fig:Running time of different algorithms in synthetic datasets} \vspace*{-0.8cm}
%\end{figure}

%\vspace*{-0.1cm}
\subsection{Memory Usage}
\vspace*{-0.1cm}

In this experiment, we compare the memory usage of the three algorithms and the size of the original dataset used. We set $t=3$, $k=2$ for Slashdot, Eponions and Super. Due to the large size of the WiKi, we set $t=8$, $t=2$. Figure \ref{fig:Memory Usage} shows the results. 

The memory usage of the three algorithms is very small, about only six times the size of the dataset. The memory usage of the three algorithms is close. It shows that our optimizations do not result in significant changes in memory usage.

\vspace*{-0.1cm}
\subsection{Case study on Wiki-RfA}
\vspace*{-0.1cm}

%\begin{figure}[t!]\vspace*{-0cm}
%	\centering
%	\includegraphics[height=5cm]{figure/case}
%	\vspace*{-0.4cm} \caption{Antagonistic $k$-plex and balanced clique}
%	\label{fig:case} 
%\end{figure}

\begin{figure}[t!] \vspace*{-0cm}
\centering
\subfigcapskip=-7pt
\subfigure[{\scriptsize Balanced clique}]{
	\includegraphics[height=4.2cm]
	{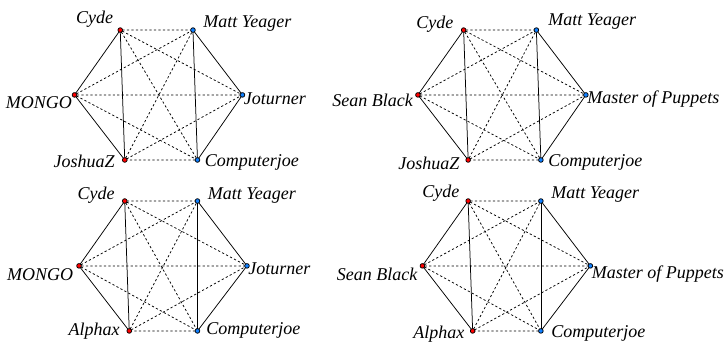}
}
\subfigure[{\scriptsize Antagonistic $k$-plex}]{
	\includegraphics[width=6cm]{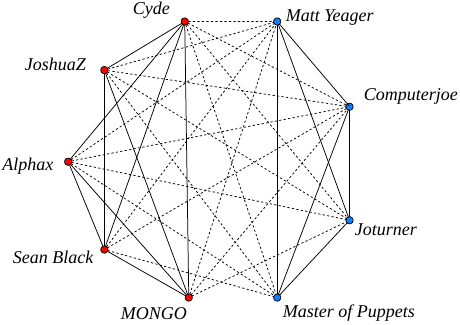}
}
\vspace*{-0.3cm}
\caption{Case study}
\label{fig:case} 
\end{figure}

In this experiment, we conduct a case study of the real dataset Wiki-RfA. Wiki-RfA records the mutual voting results of Wikipedia editors running for administrator, downloaded from https://snap.stanford.edu/. We build an undirected signed graph with wiki-RfA. Each editor can be seen as a vertex. If there is a voting relationship between two editors, then there is an edge between the two editors. If there is only support between two editors, the two editors are on a positive edge. Otherwise, they are on a negative edge. Our goal is to find cohesive antagonistic communities in the graph.

Figure \ref{fig:case} shows an example of search results for   the balanced clique model \cite{chen2020efficient} and our model. The red and blue vertices denote the two antagonistic groups in a cohesive antagonistic community. As shown in Figure \ref{fig:case}a,
\cite{chen2020efficient} can only find several 3-balanced cliques in this cohesive antagonistic community, which share many vertices. In fact, all the four 3-balanced cliques should be included in a big cohesive antagonistic community. Figure \ref{fig:case}b shows the result of our model. For antagonistic $k$-plex, we set $t=4$ and $k=2$. Obviously the antagonistic $k$-plex gives better results. All the balanced cliques in Figure \ref{fig:case}a can be included in the antagonistic $k$-plex shown in \ref{fig:case}b.

This case verifies that the \mokplex enumeration can be applied to find more generalized antagonistic communities than balanced cliques.
\section{RELATED WORKS}
%\vspace*{-0.1cm}

\stitle{Signed network.} A lot of literature on signed graph analysis has appeared in recent years. An excellent survey on signed network
analysis can be found in \cite{girdhar2017signed}. Among these theories, Structural balance theory is one of the fundamental theory\cite{DBLP:journals/isf/ZhengZW15} which is first introduced in \cite{heider1946attitudes} and extended to graph analysis in \cite{harary1953notion}. Our work is also based on structural balance theory. \cite{DBLP:conf/kdd/ChuWPWZC16,DBLP:journals/datamine/GaoLLP16,DBLP:conf/cikm/LoSZL11,su2017algorithm} aim to find the antagonistic communities in a signed network. The authors in \cite{DBLP:conf/kdd/ChuWPWZC16} hope to find a group of $k$ subgraphs such that the edges within each subgraph are dense and cohesive and the edges crossing different subgraphs are dense and oppositive. These works aim to find several groups of dense subgraphs, but they do not propose a clear structural definition of their model. 

There is also some research on searching cliques in signed graphs. an $(\alpha , k)$-clique model is proposed in \cite{li2018efficient}. They want to find a maximal clique $C$ in a signed graph such that the vertices in $C$ have no more than $k$ negative neighbors and no less than $\alpha k$ positive neighbors. However, this model only considers the positive and negative degree of vertices and ignores the model structure. To overcome this problem, \cite{chen2020efficient} propose the definition of maximal balanced clique and a new enumeration algorithm for the maximal balanced clique enumeration problem. Given a signed network $G$, a balanced clique is complete and can be split into two subcliques. The edges in the same subclique are positive, and the edges between the different subcliques are negative. The author in \cite{yaocomputing} extended the balanced clique definition and proposed an algorithm for the maximum balanced clique enumeration problem. However, the clique model is too strict to mine real cohesive antagonistic communities on signed graphs fully. Therefore, in this paper, we choose $k$-plex, the relaxed model of clique.

\stitle{$k$-plex model on unsigned networks.} It is well-known that the problem of enumerating all maximal $k$-plexes is NP-hard\cite{DBLP:journals/ior/BalasundaramBH11}. Most of the current algorithms for enumerating maximal clique or relaxation models of clique are based on the core of the algorithm in \cite{BronK73,DBLP:journals/siamcomp/TsukiyamaIAS77}. Many existing
algorithms for listing maximal $k$-plexes, as in \cite{DBLP:journals/jea/BentertHMMNS19,DBLP:journals/jpdc/WangCHSLPI17,DBLP:conf/pakdd/WuP07} also stem from the Bron-Kerbosch algorithm. Noted that when $k=1$, $k$-plex is a clique. In addition, many small size plexes are not applicable in the analysis of real data, so it is proposed in \cite{DBLP:conf/kdd/ConteFMPT17} to find extremely large plexes of size at least $q$. An efficient algorithm using clique and $k$-core to reduce the search space is also proposed in \cite{DBLP:conf/kdd/ConteFMPT17}. Then, \cite{conte2018d2k} further modified the pivot method, which is commonly used for clique enumeration, to further optimize the efficiency of the maximal $k$-plex algorithm. They also propose size pruning rules based on the number of common neighbors of two vertices\cite{conte2018d2k}. However, the worst-case time complexity of the previous algorithms is $O(n^{2}2^{n})$.  \cite{zhou2020enumerating} proposed a new algorithm which can lists all maximal $k$-plexes with provably worst-case running time $O(n^{2}\gamma ^{n})$ where $\gamma < 2$. They 
design a pivot heuristic that always branches on the vertex of the minimum degree in the graph. The time complexity of \cite{DBLP:conf/cikm/DaiLQLW22}'s method is similar, but its practical effect is better. \cite{wang2022listing} lists all maximal k-plexes in $O(n^{2}\gamma ^{D})$ time, where $\gamma$ is also smaller than 2, and $D$ is the degeneracy of the graph that is far less than the vertices number $n$. 
However, these algorithms are difficult to apply to signed graphs due to different structures.

%\vspace*{-0.2cm}
\section{CONCLUSIONS}
%\vspace*{-0.2cm}

In this paper, we study the \mokplex problem in signed graphs. We propose a new model named antagonistic $k$-plex in a signed graph. A new \mokplex enumeration framework is devised based on this model. Then, we design novel optimization strategies based on pivot and color bound to improve the efficiency of the enumeration algorithm. We also use early termination pruning and degree-based pruning to further reduce unnecessary iterative searches. We use the reduction in the dichromatic graph to reduce the search space in the preprocessing stage. Experimental results on real datasets demonstrate the efficiency, effectiveness, and scalability of our modified algorithms.

\bibliographystyle{IEEEtran}
\bibliography{IEEEabrv,sample}

% Generated by IEEEtran.bst, version: 1.14 (2015/08/26)
\begin{thebibliography}{10}
\providecommand{\url}[1]{#1}
\csname url@samestyle\endcsname
\providecommand{\newblock}{\relax}
\providecommand{\bibinfo}[2]{#2}
\providecommand{\BIBentrySTDinterwordspacing}{\spaceskip=0pt\relax}
\providecommand{\BIBentryALTinterwordstretchfactor}{4}
\providecommand{\BIBentryALTinterwordspacing}{\spaceskip=\fontdimen2\font plus
\BIBentryALTinterwordstretchfactor\fontdimen3\font minus
  \fontdimen4\font\relax}
\providecommand{\BIBforeignlanguage}[2]{{%
\expandafter\ifx\csname l@#1\endcsname\relax
\typeout{** WARNING: IEEEtran.bst: No hyphenation pattern has been}%
\typeout{** loaded for the language `#1'. Using the pattern for}%
\typeout{** the default language instead.}%
\else
\language=\csname l@#1\endcsname
\fi
#2}}
\providecommand{\BIBdecl}{\relax}
\BIBdecl

\bibitem{DBLP:journals/tasm/Chen13}
H.~Chen, ``Networks, crowds, and markets: Reasoning about a highly connected
  world [book review],'' \emph{{IEEE} Technol. Soc. Mag.}, vol.~32, no.~3,
  p.~10, 2013.

\bibitem{DBLP:conf/www/KunegisLB09}
J.~Kunegis, A.~Lommatzsch, and C.~Bauckhage, ``The slashdot zoo: mining a
  social network with negative edges,'' in \emph{Proceedings of the 18th
  International Conference on World Wide Web, {WWW} 2009, Madrid, Spain, April
  20-24, 2009}, J.~Quemada, G.~Le{\'{o}}n, Y.~S. Maarek, and W.~Nejdl,
  Eds.\hskip 1em plus 0.5em minus 0.4em\relax {ACM}, 2009, pp. 741--750.

\bibitem{DBLP:conf/sdm/GiatsidisCMTV14}
C.~Giatsidis, B.~Cautis, S.~Maniu, D.~M. Thilikos, and M.~Vazirgiannis,
  ``Quantifying trust dynamics in signed graphs, the s-cores approach,'' in
  \emph{Proceedings of the 2014 {SIAM} International Conference on Data Mining,
  Philadelphia, Pennsylvania, USA, April 24-26, 2014}, M.~J. Zaki,
  Z.~Obradovic, P.~Tan, A.~Banerjee, C.~Kamath, and S.~Parthasarathy,
  Eds.\hskip 1em plus 0.5em minus 0.4em\relax {SIAM}, 2014, pp. 668--676.

\bibitem{DBLP:journals/tcbb/Ou-YangDZ15}
L.~Ou{-}Yang, D.~Dai, and X.~Zhang, ``Detecting protein complexes from signed
  protein-protein interaction networks,'' \emph{{IEEE} {ACM} Trans. Comput.
  Biol. Bioinform.}, vol.~12, no.~6, pp. 1333--1344, 2015.

\bibitem{harary1953notion}
F.~Harary, ``On the notion of balance of a signed graph.'' \emph{Michigan
  Mathematical Journal}, vol.~2, no.~2, pp. 143--146, 1953.

\bibitem{DBLP:journals/datamine/GaoLLP16}
M.~Gao, E.~Lim, D.~Lo, and P.~K. Prasetyo, ``On detecting maximal quasi
  antagonistic communities in signed graphs,'' \emph{Data Min. Knowl. Discov.},
  vol.~30, no.~1, pp. 99--146, 2016.

\bibitem{chen2020efficient}
Z.~Chen, L.~Yuan, X.~Lin, L.~Qin, and J.~Yang, ``Efficient maximal balanced
  clique enumeration in signed networks,'' in \emph{Proceedings of The Web
  Conference 2020}, 2020, pp. 339--349.

\bibitem{yaocomputing}
K.~Yao, L.~Chang, and L.~Qin, ``Computing maximum structural balanced cliques
  in signed graphs,'' in \emph{38th {IEEE} International Conference on Data
  Engineering, {ICDE} 2022, Kuala Lumpur, Malaysia, May 9-12, 2022}.\hskip 1em
  plus 0.5em minus 0.4em\relax {IEEE}, 2022, pp. 1004--1016.

\bibitem{DBLP:conf/kdd/ConteFMPT17}
A.~Conte, D.~Firmani, C.~Mordente, M.~Patrignani, and R.~Torlone, ``Fast
  enumeration of large k-plexes,'' in \emph{Proceedings of the 23rd {ACM}
  {SIGKDD} International Conference on Knowledge Discovery and Data Mining,
  Halifax, NS, Canada, August 13 - 17, 2017}.\hskip 1em plus 0.5em minus
  0.4em\relax {ACM}, 2017, pp. 115--124.

\bibitem{DBLP:journals/ior/BalasundaramBH11}
B.~Balasundaram, S.~Butenko, and I.~V. Hicks, ``Clique relaxations in social
  network analysis: The maximum \emph{k}-plex problem,'' \emph{Oper. Res.},
  vol.~59, no.~1, pp. 133--142, 2011.

\bibitem{kumar2018community}
S.~Kumar, W.~L. Hamilton, J.~Leskovec, and D.~Jurafsky, ``Community interaction
  and conflict on the web,'' in \emph{Proceedings of the 2018 world wide web
  conference}, 2018, pp. 933--943.

\bibitem{xiao2020searching}
H.~Xiao, B.~Ordozgoiti, and A.~Gionis, ``Searching for polarization in signed
  graphs: a local spectral approach,'' in \emph{Proceedings of The Web
  Conference 2020}, 2020, pp. 362--372.

\bibitem{bonchi2019discovering}
F.~Bonchi, E.~Galimberti, A.~Gionis, B.~Ordozgoiti, and G.~Ruffo, ``Discovering
  polarized communities in signed networks,'' in \emph{Proceedings of the 28th
  acm international conference on information and knowledge management}, 2019,
  pp. 961--970.

\bibitem{suratanee2014characterizing}
A.~Suratanee, M.~H. Schaefer, M.~J. Betts, Z.~Soons, H.~Mannsperger, N.~Harder,
  M.~Oswald, M.~Gipp, E.~Ramminger, G.~Marcus \emph{et~al.}, ``Characterizing
  protein interactions employing a genome-wide sirna cellular phenotyping
  screen,'' \emph{PLoS computational biology}, vol.~10, no.~9, p. e1003814,
  2014.

\bibitem{yim2018annotating}
S.~Yim, H.~Yu, D.~Jang, and D.~Lee, ``Annotating activation/inhibition
  relationships to protein-protein interactions using gene ontology
  relations,'' \emph{BMC systems biology}, vol.~12, no.~1, pp. 111--122, 2018.

\bibitem{ou2015detecting}
L.~Ou-Yang, D.-Q. Dai, and X.-F. Zhang, ``Detecting protein complexes from
  signed protein-protein interaction networks,'' \emph{IEEE/ACM transactions on
  computational biology and bioinformatics}, vol.~12, no.~6, pp. 1333--1344,
  2015.

\bibitem{miller1995wordnet}
G.~A. Miller, ``Wordnet: a lexical database for english,'' \emph{Communications
  of the ACM}, vol.~38, no.~11, pp. 39--41, 1995.

\bibitem{kumar2019cross}
V.~Kumar, N.~Joshi, A.~Mukherjee, G.~Ramakrishnan, and P.~Jyothi,
  ``Cross-lingual training for automatic question generation,'' \emph{arXiv
  preprint arXiv:1906.02525}, 2019.

\bibitem{krishnan2018leveraging}
A.~Krishnan, D.~P, S.~Ranu, and S.~Mehta, ``Leveraging semantic resources in
  diversified query expansion,'' \emph{World Wide Web}, vol.~21, pp.
  1041--1067, 2018.

\bibitem{seidman1978graph}
S.~B. Seidman and B.~L. Foster, ``A graph-theoretic generalization of the
  clique concept,'' \emph{Journal of Mathematical sociology}, vol.~6, no.~1,
  pp. 139--154, 1978.

\bibitem{DBLP:conf/aaai/ZhouXGXJ20}
Y.~Zhou, J.~Xu, Z.~Guo, M.~Xiao, and Y.~Jin, ``Enumerating maximal
  \emph{k}-plexes with worst-case time guarantee,'' in \emph{The Thirty-Fourth
  {AAAI} Conference on Artificial Intelligence, {AAAI} 2020, The Thirty-Second
  Innovative Applications of Artificial Intelligence Conference, {IAAI} 2020,
  The Tenth {AAAI} Symposium on Educational Advances in Artificial
  Intelligence, {EAAI} 2020, New York, NY, USA, February 7-12, 2020}.\hskip 1em
  plus 0.5em minus 0.4em\relax {AAAI} Press, 2020, pp. 2442--2449.

\bibitem{conte2018d2k}
A.~Conte, T.~De~Matteis, D.~De~Sensi, R.~Grossi, A.~Marino, and L.~Versari,
  ``D2k: scalable community detection in massive networks via small-diameter
  k-plexes,'' in \emph{Proceedings of the 24th ACM SIGKDD International
  Conference on Knowledge Discovery \& Data Mining}, 2018, pp. 1272--1281.

\bibitem{balasundaram2011clique}
B.~Balasundaram, S.~Butenko, and I.~V. Hicks, ``Clique relaxations in social
  network analysis: The maximum k-plex problem,'' \emph{Operations Research},
  vol.~59, no.~1, pp. 133--142, 2011.

\bibitem{wang2022listing}
Z.~Wang, Y.~Zhou, M.~Xiao, and B.~Khoussainov, ``Listing maximal k-plexes in
  large real-world graphs,'' in \emph{Proceedings of the ACM Web Conference
  2022}, 2022, pp. 1517--1527.

\bibitem{wang2017parallelizing}
Z.~Wang, Q.~Chen, B.~Hou, B.~Suo, Z.~Li, W.~Pan, and Z.~G. Ives,
  ``Parallelizing maximal clique and k-plex enumeration over graph data,''
  \emph{Journal of Parallel and Distributed Computing}, vol. 106, pp. 79--91,
  2017.

\bibitem{wu2007parallel}
B.~Wu and X.~Pei, ``A parallel algorithm for enumerating all the maximal
  k-plexes,'' in \emph{Pacific-Asia conference on knowledge discovery and data
  mining}.\hskip 1em plus 0.5em minus 0.4em\relax Springer, 2007, pp. 476--483.

\bibitem{zhou2021improving}
Y.~Zhou, S.~Hu, M.~Xiao, and Z.-H. Fu, ``Improving maximum k-plex solver via
  second-order reduction and graph color bounding,'' in \emph{Proceedings of
  the AAAI Conference on Artificial Intelligence}, vol.~35, no.~14, 2021, pp.
  12\,453--12\,460.

\bibitem{zhou2020enumerating}
Y.~Zhou, J.~Xu, Z.~Guo, M.~Xiao, and Y.~Jin, ``Enumerating maximal k-plexes
  with worst-case time guarantee,'' in \emph{Proceedings of the AAAI Conference
  on Artificial Intelligence}, vol.~34, no.~03, 2020, pp. 2442--2449.

\bibitem{li2018efficient}
R.-H. Li, Q.~Dai, L.~Qin, G.~Wang, X.~Xiao, J.~X. Yu, and S.~Qiao, ``Efficient
  signed clique search in signed networks,'' in \emph{2018 IEEE 34th
  International Conference on Data Engineering (ICDE)}.\hskip 1em plus 0.5em
  minus 0.4em\relax IEEE, 2018, pp. 245--256.

\bibitem{girdhar2017signed}
N.~Girdhar and K.~Bharadwaj, ``Signed social networks: a survey,'' in
  \emph{Advances in Computing and Data Sciences: First International
  Conference, ICACDS 2016, Ghaziabad, India, November 11-12, 2016, Revised
  Selected Papers 1}.\hskip 1em plus 0.5em minus 0.4em\relax Springer, 2017,
  pp. 326--335.

\bibitem{DBLP:journals/isf/ZhengZW15}
X.~Zheng, D.~D. Zeng, and F.~Wang, ``Social balance in signed networks,''
  \emph{Inf. Syst. Frontiers}, vol.~17, no.~5, pp. 1077--1095, 2015.

\bibitem{heider1946attitudes}
F.~Heider, ``Attitudes and cognitive organization,'' \emph{The Journal of
  psychology}, vol.~21, no.~1, pp. 107--112, 1946.

\bibitem{DBLP:conf/kdd/ChuWPWZC16}
L.~Chu, Z.~Wang, J.~Pei, J.~Wang, Z.~Zhao, and E.~Chen, ``Finding gangs in war
  from signed networks,'' in \emph{Proceedings of the 22nd {ACM} {SIGKDD}
  International Conference on Knowledge Discovery and Data Mining, San
  Francisco, CA, USA, August 13-17, 2016}, B.~Krishnapuram, M.~Shah, A.~J.
  Smola, C.~C. Aggarwal, D.~Shen, and R.~Rastogi, Eds.\hskip 1em plus 0.5em
  minus 0.4em\relax {ACM}, 2016, pp. 1505--1514.

\bibitem{DBLP:conf/cikm/LoSZL11}
D.~Lo, D.~Surian, K.~Zhang, and E.~Lim, ``Mining direct antagonistic
  communities in explicit trust networks,'' in \emph{Proceedings of the 20th
  {ACM} Conference on Information and Knowledge Management, {CIKM} 2011,
  Glasgow, United Kingdom, October 24-28, 2011}, C.~Macdonald, I.~Ounis, and
  I.~Ruthven, Eds.\hskip 1em plus 0.5em minus 0.4em\relax {ACM}, 2011, pp.
  1013--1018.

\bibitem{su2017algorithm}
Y.~Su, B.~Wang, F.~Cheng, L.~Zhang, X.~Zhang, and L.~Pan, ``An algorithm based
  on positive and negative links for community detection in signed networks,''
  \emph{Scientific reports}, vol.~7, no.~1, pp. 1--12, 2017.

\bibitem{BronK73}
C.~Bron and J.~Kerbosch, ``Finding all cliques of an undirected graph
  (algorithm 457),'' \emph{Commun. {ACM}}, vol.~16, no.~9, pp. 575--576, 1973.

\bibitem{DBLP:journals/siamcomp/TsukiyamaIAS77}
S.~Tsukiyama, M.~Ide, H.~Ariyoshi, and I.~Shirakawa, ``A new algorithm for
  generating all the maximal independent sets,'' \emph{{SIAM} J. Comput.},
  vol.~6, no.~3, pp. 505--517, 1977.

\bibitem{DBLP:journals/jea/BentertHMMNS19}
M.~Bentert, A.~Himmel, H.~Molter, M.~Morik, R.~Niedermeier, and
  R.~Saitenmacher, ``Listing all maximal \emph{k}-plexes in temporal graphs,''
  \emph{{ACM} J. Exp. Algorithmics}, vol.~24, no.~1, pp. 1.13:1--1.13:27, 2019.

\bibitem{DBLP:journals/jpdc/WangCHSLPI17}
Z.~Wang, Q.~Chen, B.~Hou, B.~Suo, Z.~Li, W.~Pan, and Z.~G. Ives,
  ``Parallelizing maximal clique and k-plex enumeration over graph data,''
  \emph{J. Parallel Distributed Comput.}, vol. 106, pp. 79--91, 2017.

\bibitem{DBLP:conf/pakdd/WuP07}
B.~Wu and X.~Pei, ``A parallel algorithm for enumerating all the maximal
  \emph{k} -plexes,'' in \emph{Emerging Technologies in Knowledge Discovery and
  Data Mining, {PAKDD} 2007, International Workshops, Nanjing, China, May
  22-25, 2007, Revised Selected Papers}, ser. Lecture Notes in Computer
  Science, T.~Washio, Z.~Zhou, J.~Z. Huang, X.~Hu, J.~Li, C.~Xie, J.~He,
  D.~Zou, K.~Li, and M.~M. Freire, Eds., vol. 4819.\hskip 1em plus 0.5em minus
  0.4em\relax Springer, 2007, pp. 476--483.

\bibitem{DBLP:conf/cikm/DaiLQLW22}
Q.~Dai, R.~Li, H.~Qin, M.~Liao, and G.~Wang, ``Scaling up maximal \emph{k}-plex
  enumeration,'' in \emph{Proceedings of the 31st {ACM} International
  Conference on Information {\&} Knowledge Management, Atlanta, GA, USA,
  October 17-21, 2022}, M.~A. Hasan and L.~Xiong, Eds.\hskip 1em plus 0.5em
  minus 0.4em\relax {ACM}, 2022, pp. 345--354.

\end{thebibliography}

\vfill

\end{document}